\def\longonly#1{#1}
\def\shortonly#1{}
\def\nonblind#1{#1}
\def\onlyblind#1{}

\documentclass{article}
\usepackage{amsthm}
\usepackage{amsmath}
\usepackage{amssymb}
\usepackage{graphicx}
\usepackage{hyperref}

\newtheorem{theorem}{Theorem}
\newtheorem{definition}[theorem]{Definition}

\newtheorem{property}[theorem]{Property}

\def\subfig#1#2#3{
\begin{minipage}[b]{#1} \noindent
  \includegraphics[width=\textwidth]{#2}
  \begin{center}
  \vskip -1em
  #3
  \end{center}
\end{minipage}
}

\def\bbN{\mathbb{N}}
\def\bbR{\mathbb{R}}
\def\bbH{\mathbb{H}}

\def\ra{\rightarrow}

\def\calS{\mathcal{S}}

\def\InitCounter{\textsc{InitCounter}}

\def\Add{\textsc{Add}}

\def\Count{\textsc{Count}}
\def\val{{\rm val}}
\def\restto#1{|_{#1}}

\def\bbN{{\mathbb N}}
\def\bbH{{\mathbb H}}
\def\bbR{{\mathbb R}}

\def\ra{\rightarrow}

\def\dist{\delta}

\def\ext#1{#1}

\def\gshort#1#2#3{G_{#1#2#3}}

\def\ghoz{\gshort 710}
\def\ghoo{\gshort 711}
\def\gooz{\gshort 810}
\def\gqab{\gshort qab}

\title{Dynamic Distances in Hyperbolic Graphs}

\author {
    Eryk Kopczyński, Dorota Celińska-Kopczyńska \\
    { \small Institute of Informatics, University of Warsaw}
}

\begin{document}


\maketitle

\begin{abstract}
We consider the following dynamic problem: given a fixed (small) template graph with colored vertices $C$ and a large graph with colored vertices $G$ (whose colors
can be changed dynamically), how many mappings
$m$ are there from the vertices of $C$ to vertices of $G$ in such a way that the colors agree, and the distances between $m(v)$ and $m(w)$ have given
values for every edge? We show that this problem can be solved efficiently on triangulations of the hyperbolic plane, as well as other Gromov
hyperbolic graphs. For various template graphs $C$, this result lets us 
efficiently solve various computational problems which are relevant in applications, such as visualization of hierarchical data and social network analysis.
\end{abstract}

\section{Introduction}
Consider a metric space $(X,\delta)$. We would like to answer questions such as the following:

\begin{itemize}
\item Let $A$ be a large finite subset of $X$. What is the average $\delta(a_1,a_2)$ for $a_1, a_2 \in A$?
\item For $A$ as above, what is the average $\delta(a,x)$ for $a \in A$ and $x \in X$?
\item For $A$ and $a$ as above, what is the number of pairs of vertices $a_1,a_2 \in A$ such that $a$ is on a shortest path from
$a_1$ to $a_2$, i.e., $\delta(a_1,a) + \delta(a,a_2) = \delta(a_1, a_2)$?
\item For $A$ as above, consider the graph $(A,E)$, where every pair of vertices $a_1,a_2\in A$ is connected with an edge with 
probability $p(\delta(a_1,a_2))$. What is the expected average degree and the expected number of triangles in such a graph?
\end{itemize}

The graph $(A,E)$ (for a random $A$) obtained above is called a random geometric graph. Random geometric graphs are used in social network analysis,
as they exhibit the community structure typical to real-life networks. While traditionally $X$ was taken to be a bounded
subset of an Euclidean space, recently models based on hyperbolic geometry have gained popularity among big data analysts. Hyperbolic spaces have
tree-like structure, with exponentially many vertices in given distance $R$ from $v_0$; this property makes them useful
in the visualization \cite{munzner,lampingrao} and modeling of hierarchical data.
Graphs generated according to the Hyperbolic Random Graph (HRG) model
have properties (such as degree distribution and clustering coefficient) similar to that of real-world scale-free
networks \cite{papa}. Efficiently solving computational problems similar to the ones listed above is crucial when working with the HRG model.

In this paper, we present a unified framework for efficiently solving such problems,
assuming that $X$ is a disk of diameter $R$ in a fixed regularly generated triangulation of the hyperbolic plane, or in general, a Gromov hyperbolic graph of a fixed diameter,
degree and Gromov hyperbolicity (the size of such graphs can be exponential in $R$).
Gromov hyperbolicity \cite{Gromov1987}
of a graph $G$ measures whether the shortest paths in $G$ behave in a tree-like way.
In a tree, the shortest path from $a$ to $c$ is always a subset of $U$, the union of the shortest path from $a$ to $b$,
and the shortest path from $b$ to $c$. In a graph of Gromov hyperbolicity $\delta$, 
(any) shortest path from $a$ to $c$ is always in the $\delta$-neighborhood of $U$.
Our framework generalizes the theoretical ideas underlying our another paper \cite{dhrgex},
which focuses on the experimental results of applying them to the HRG model.

In our framework, we fix a set of colors $K$, and a template colored graph $C = (V_C, E_C, k:V_C\ra K)$. We can dynamically assign colors from 
$K$ to points in $X$, and ask queries of the following form: 
``how many embeddings $m:V_C \ra X$ are there such that every $v\in V_C$ is mapped to a vertex
of color $k(v)$, and distances from $m(a)$ to $m(b)$ for every $(a,b) \in E_C$ are given?''.
We show that we can recolor the (lazily generated) vertices of $G$
in time $O(R^{|V_C|+|E_C|})$ and reanswer the question in $O(1)$.
Note that this time is independent from the number of vertices colored so far, which can be exponential in $R$,
and in fact, in most of our applications, $R$ can be considered logarithmic in the number of vertices colored.
Thus, for example, the average distance above can be solved by taking $C$ consisting of two vertices of the same color $k_1$,
and edge between them. After coloring every point in $A$ with color $k_1$, we can compute the number of pairs of points in
every distance from $0$ to $R$, and thus compute the average distance. This algorithm runs in time $O(nR^3)$; in this specific
case we can actually achieve $O(R^2)$ update time and thus $O(nR^2)$ total running time, which in our application is significantly
better than the trivial $O(n^2)$ algorithm. The second average distance problem can be solved similarly -- we instead change the
color of one of the vertices in $C$ to $k_2$, and color $x$ with $k_2$.

Tessellations of the hyperbolic plane are useful in visualization \cite{hrviz}, 
dimension reduction algorithms \cite{ritter99,ontrup} and video game design \cite{hyperrogue,hypminesweeper}. While the HRG model
traditionally uses the hyperbolic plane in its continuous form, using a discrete triangulation is a promising approach,
as it lets us to avoiding precision issues inherent to coordinate-based models of hyperbolic geometry \cite{tobias_alenex}. 

Our main result takes inspiration from the Courcelle's theorem \cite{courcelle}. 
It is well known in theoretical computer science that many 
computational problems admit efficient solutions on trees.
Usually, these solutions involve running a dynamic programming algorithm
over the tree. Courcelle's theorem says that,
for any fixed $d$ and any fixed formula $\phi$ of Monadic Second Order logic
with quantification over sets of vertices and edges (MSO$_2$), 
it can be verified whether the given graph $G=(X,E)$ of treewidth bounded by $d$ 
satisfies the formula $\phi$ in linear time. Courcelle's theorem gives a general method 
of constructing efficient algorithms working on graphs similar to trees, where the similarity to a
tree is measured using the {\it treewidth} parameter. 
Our result is different, because it is not based on bounded treewidth;
while it is common for 
graphs naturally embeddable in $\bbH^2$ to have bounded treewidth \cite{hyptreewidth,hypminesweeper}, this is no
longer the case in higher dimensions: a graph similar to $\bbH^3$ may contain a Euclidean two-dimensional grid,
which has a very large treewidth, and is an obstacle for efficient model checking of formulas in logic similar to MSO.
The notion of tree-likeness appropriate for us is Gromov hyperbolicity, and instead of logical formulas, we use
a template graph which specifies the configuration of distances we are looking for.

\paragraph*{Structure of the paper}
In Section \ref{sec:htgrid}, we prepare the ground for dealing algorithmically with regular and Goldberg-Coxeter triangulations of the
hyperbolic plane. Such triangulations are typical examples of graphs embeddable in the hyperbolic plane.
Up to our knowledge, algorithms for dealing with such triangulations were not previously explored in as much detail.
In Section \ref{sec:stg}, we define segment tree graphs that generalize triangulations from Section \ref{sec:htgrid} and prove our main result. Segment tree graphs are exponentially expanding
graphs that behave similar to tessellations of hyperbolic spaces. 
Section \ref{sec:other} generalizes our results to all Gromov hyperbolic graphs of bounded degree.
Section \ref{sec:apps} discusses the applications.

\nonblind{
\paragraph*{Acknowledgments}
This work has been supported by the National Science Centre, Poland, grant UMO-2019//35/B/ST6/04456.
}

\section{Hyperbolic triangulations}\label{sec:htgrid}

\def\sch#1#2{\{#1,#2\}}
\def\scht#1{\sch{3}{#1}}
\def\schq#1{\sch{4}{#1}}
\def\gp#1#2{GC_{#1,#2}}

\begin{figure}[ht]
\begin{center}
\longonly{
\subfig{0.4\linewidth}{img/tiling-hep.pdf}{(a)}
\subfig{0.4\linewidth}{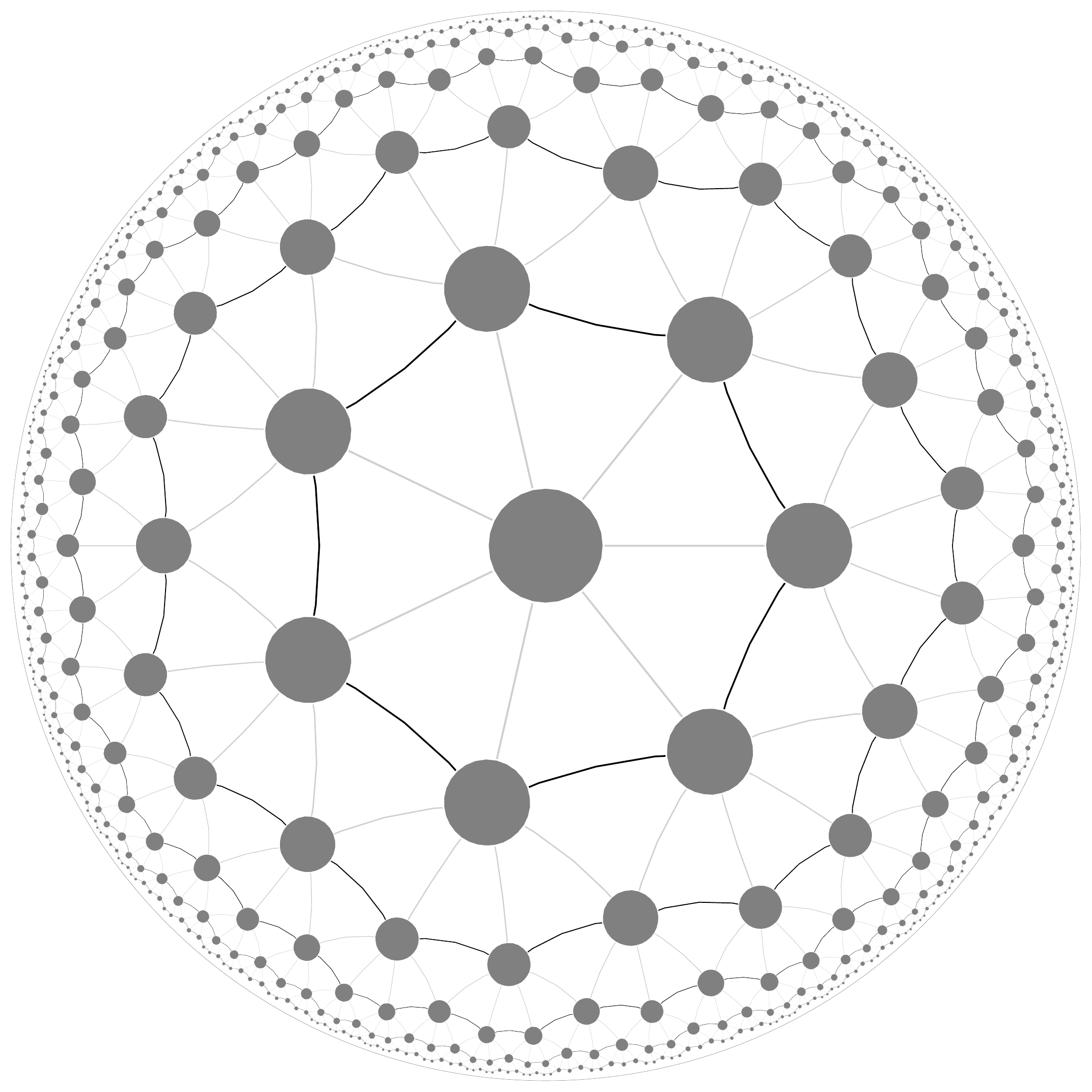}{(b)}
\subfig{0.4\linewidth}{img/tiling-hr.pdf}{(c)}
\subfig{0.4\linewidth}{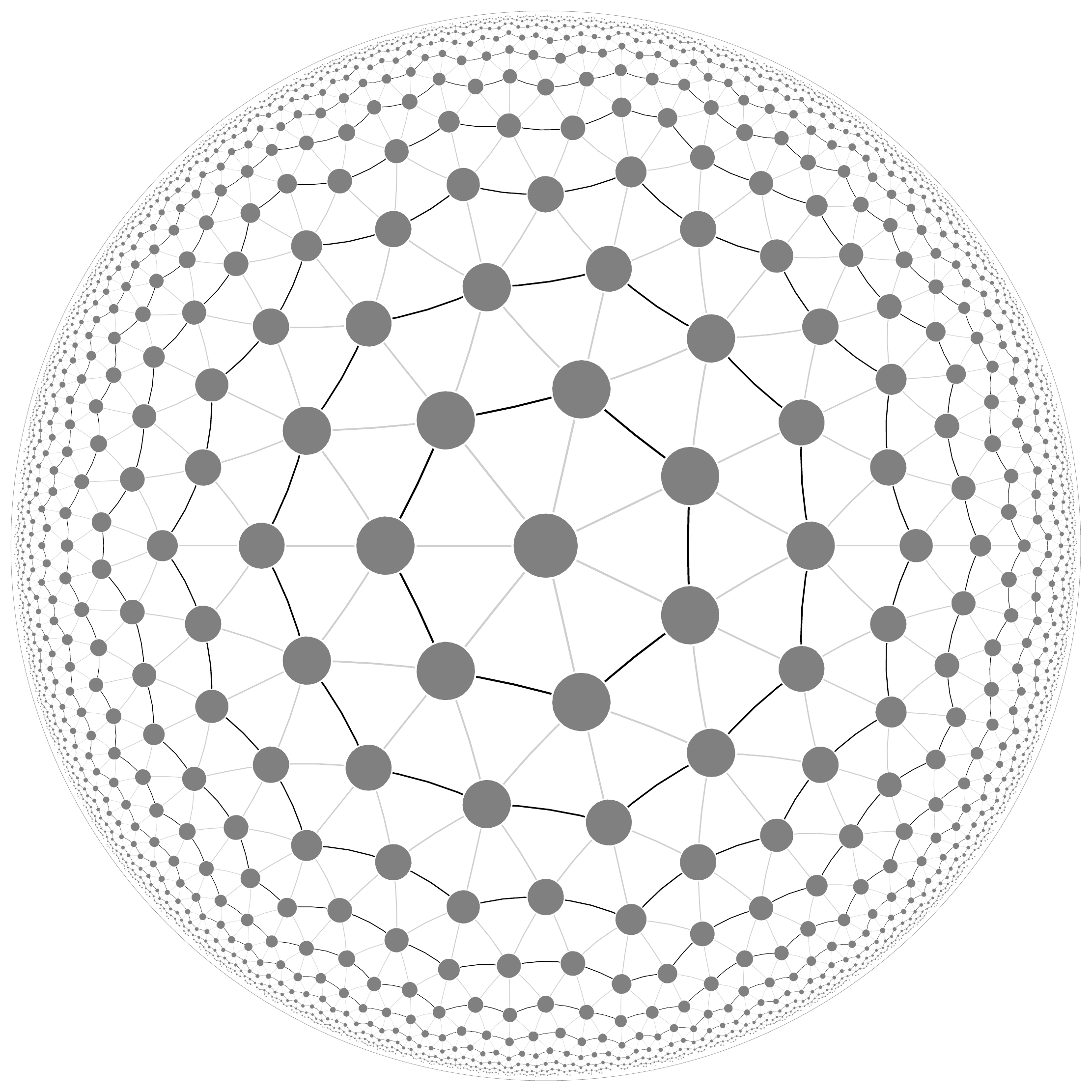}{(d)}
}
\shortonly{
\subfig{0.2\linewidth}{img/tiling-hep.pdf}{(a)}
\subfig{0.2\linewidth}{img/trigrid7.pdf}{(b)}
\subfig{0.2\linewidth}{img/tiling-hr.pdf}{(c)}
\subfig{0.2\linewidth}{img/trigrid67.pdf}{(d)}
}
\end{center}
\caption{\label{figtile}
(a) order-3 heptagonal tiling, (b) the triangulation $\ghoz$,
(c) truncated triangular tiling, (d) the triangulation $\ghoo$.}
\end{figure}

Let $\bbN=\{0,1,\ldots\}$.
Figure \ref{figtile} shows two tilings of the hyperbolic plane, the order-3 heptagonal
tiling and its bitruncated variant, in the Poincar\'e disk model, together
with their dual graphs, which we call $\ghoz$ and $\ghoo$. In the Poincar\'e
model, the hyperbolic plane is represented as a disk. 
In the hyperbolic metric, 
all the triangles, heptagons and hexagons on each of these pictures are actually of the same size,
and the points on the boundary of the disk are infinitely far from the center.%
\nonblind{\footnote{See \url{https://www.mimuw.edu.pl/~erykk/dhrg} for an interactive visualization.}}

In a regular tesselation every face is a regular $p$-gon, and every vertex has degree $q$ (we assume $p,q\geq 3$). We say that such a tesselation has a 
{\bf Schl\"afli symbol} $\sch{p}{q}$. Such a tesselation exists on the sphere if and only if $(p-2)(q-2)<4$, plane if and only if $(p-2)(q-2)=4$, and hyperbolic plane if and only if $(p-2)(q-2) > 4$.
In this paper, we are most interested in triangulations ($p=3$) of the hyperbolic plane ($q>6$).

Contrary to the Euclidean tesselations, hyperbolic tesselations cannot be scaled: on a hyperbolic plane of curvature -1, every face in a $\sch{q}{p}$ tesselation, and equivalently
the set of points closest to the given vertex in its dual $\sch{p}{q}$ tesselation, will have area $\pi(q\frac{p-2}{p}-2)$. Thus, among hyperbolic triangulations of the form $\scht{q}$, $\scht{7}$
is the finest, and they get coarser and coarser as $q$ increases.

\begin{figure}[ht]
\begin{center}
\shortonly{
\subfig{0.24\linewidth}{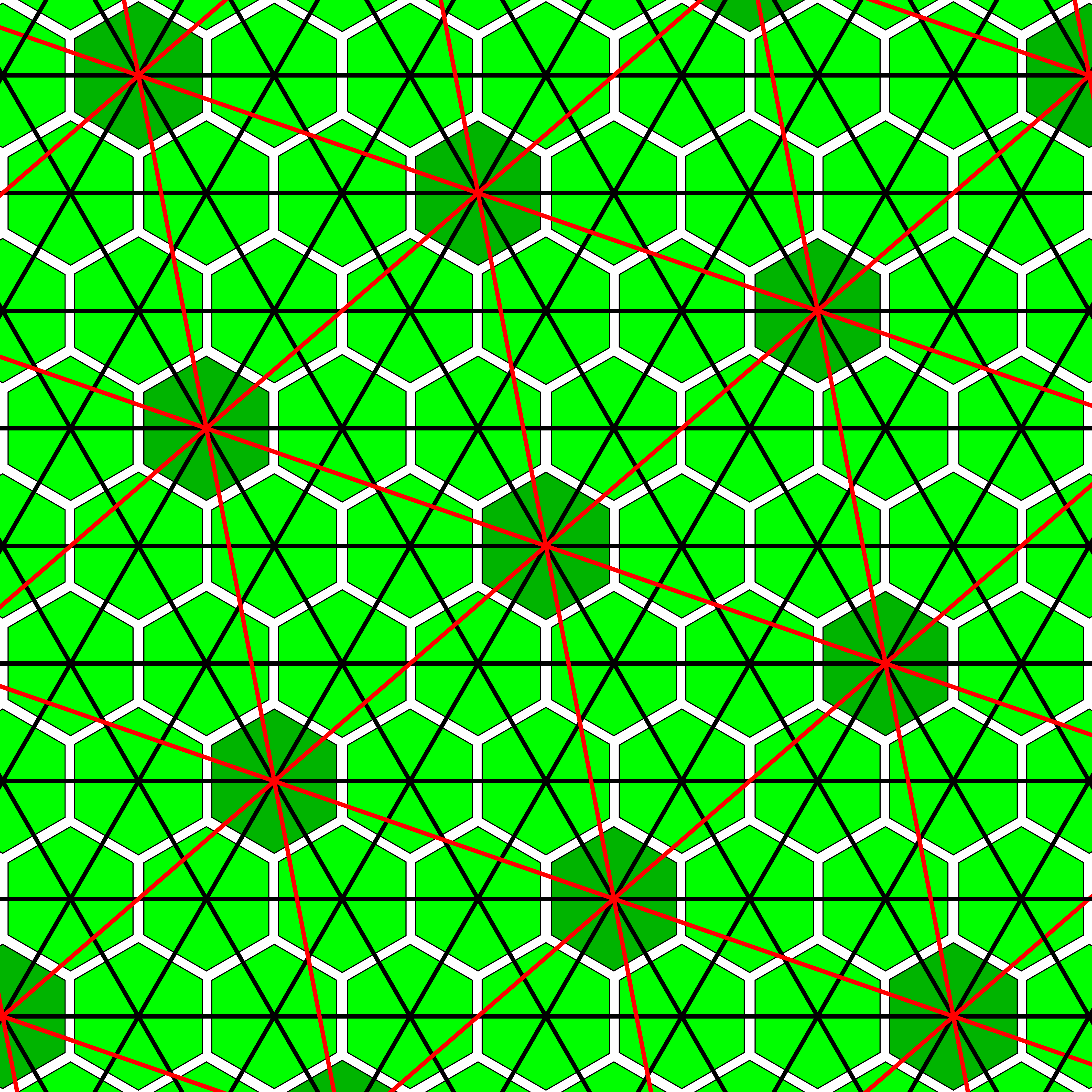}{(a) Euclidean plane}
\hskip -1mm
\subfig{0.24\linewidth}{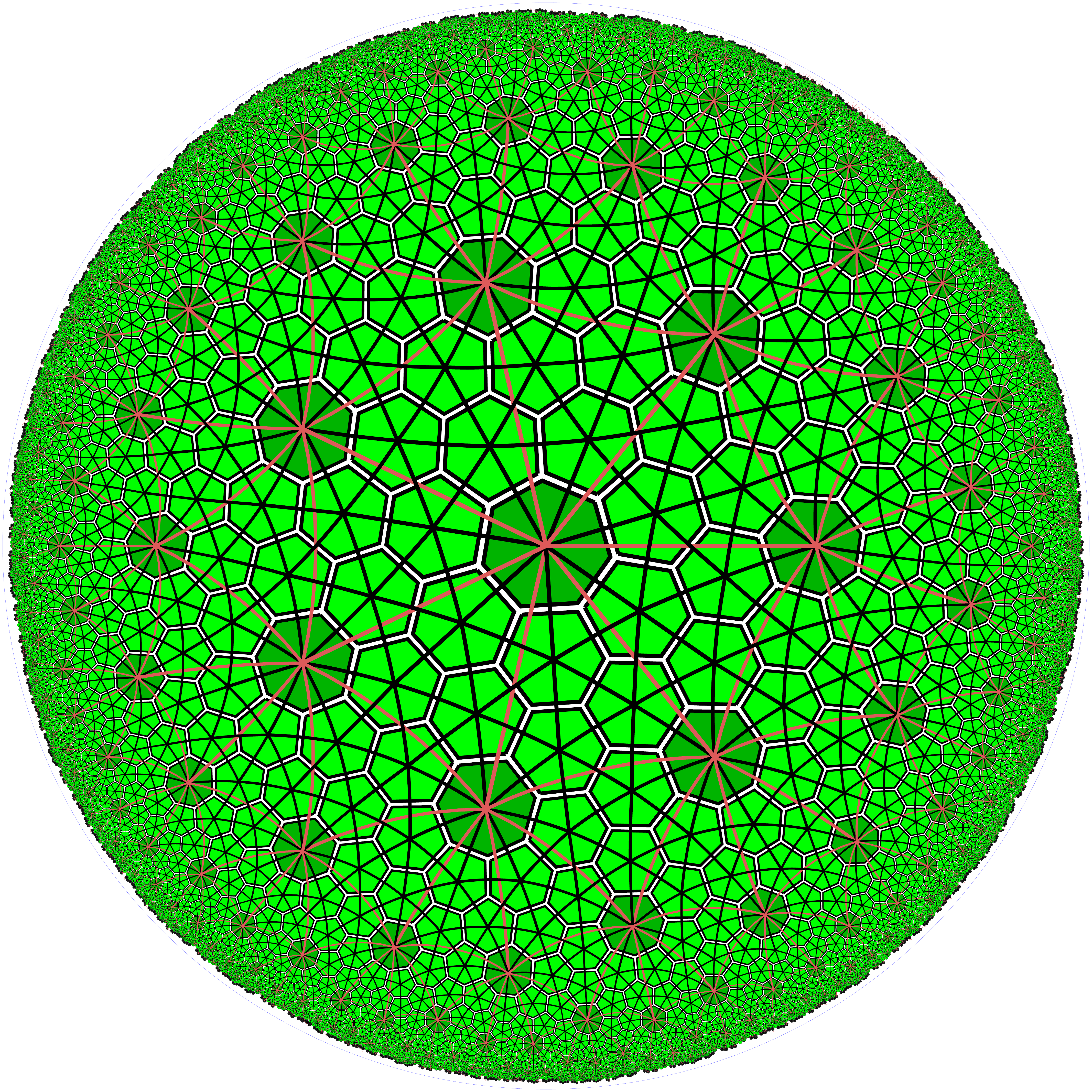}{(b) $\gshort 721$}
\hskip -1mm
\subfig{0.24\linewidth}{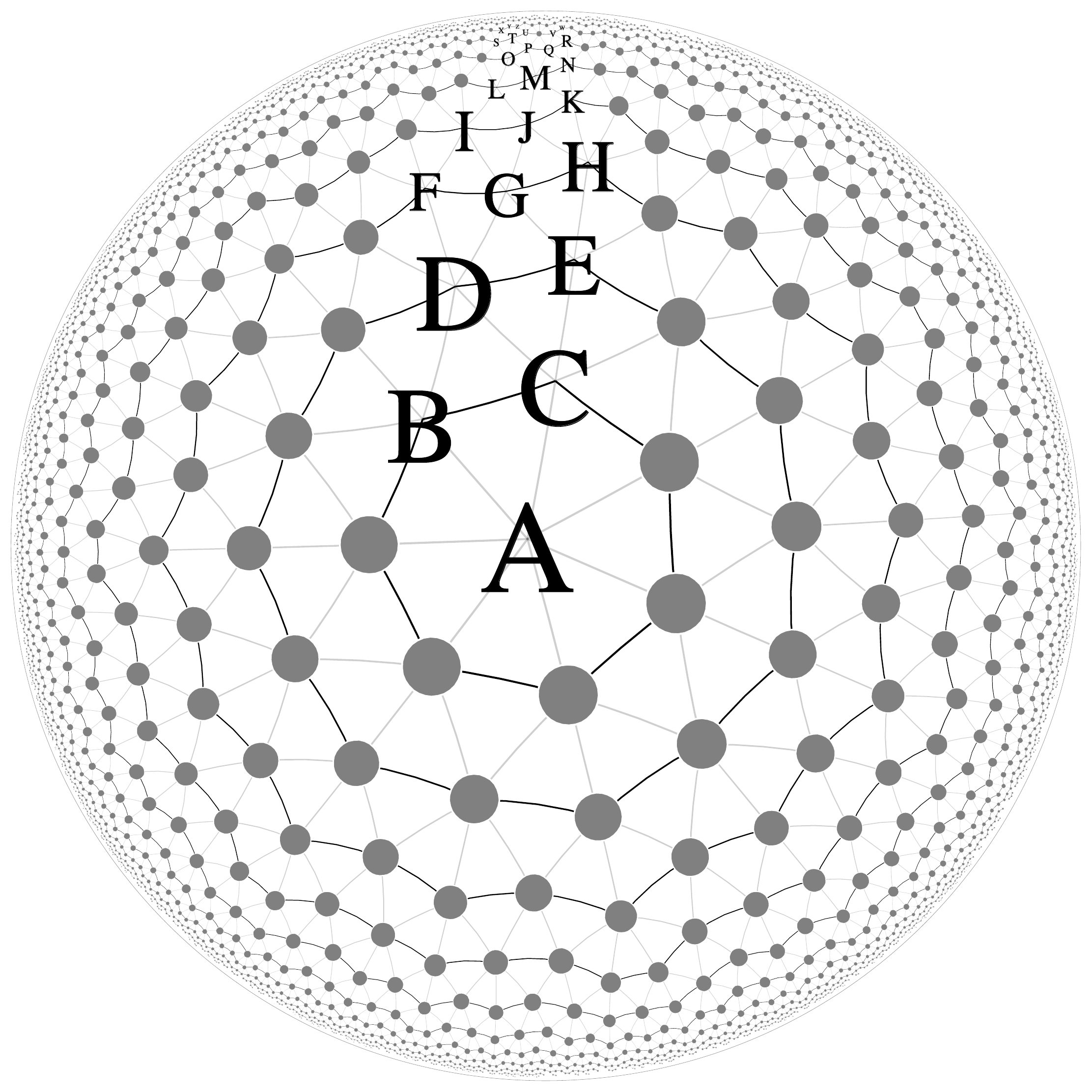}{(c) $\ghoo$}
\hskip -1mm
\subfig{0.24\linewidth}{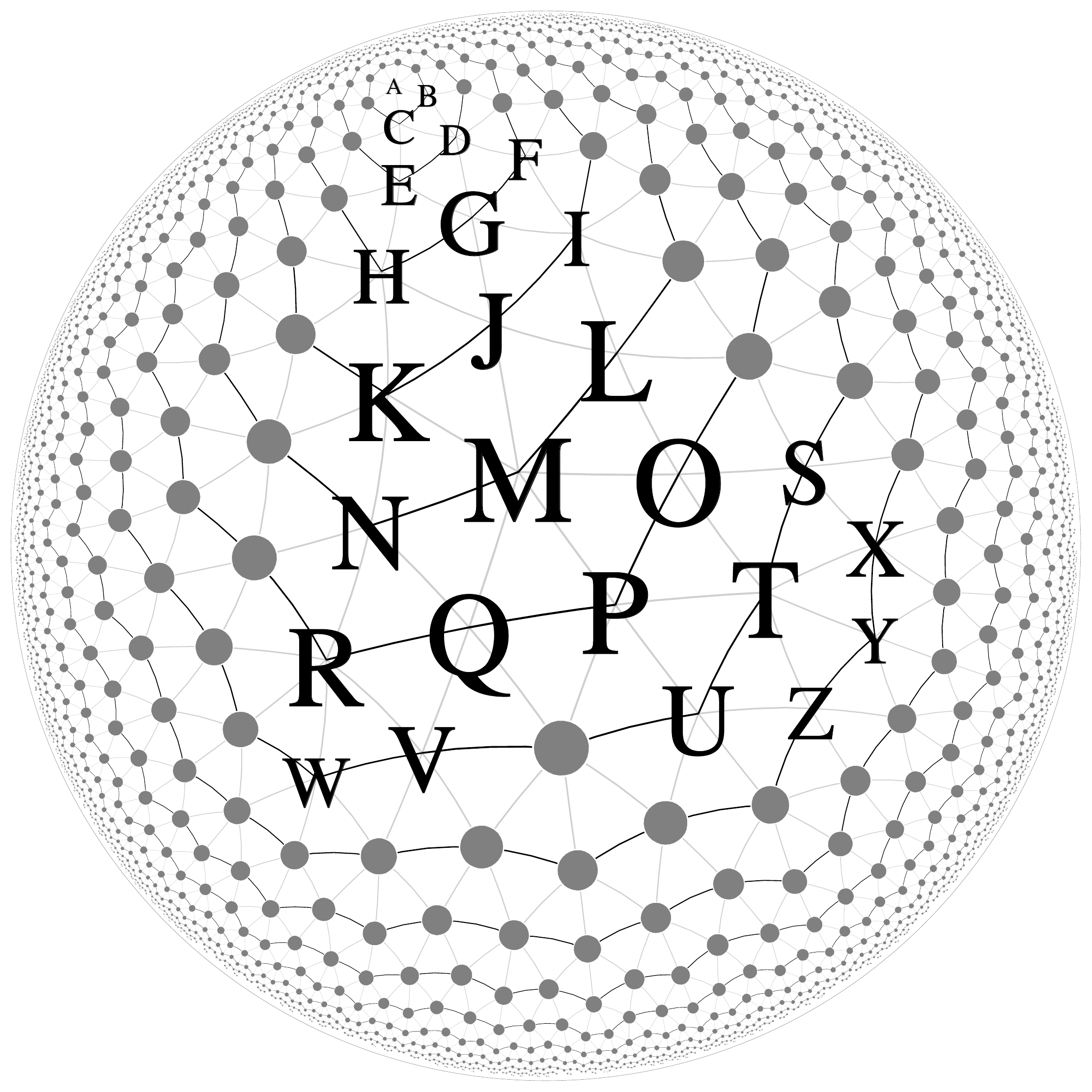}{(d) not centered}
}
\longonly{
\subfig{0.4\linewidth}{img/goldberg-eu.pdf}{(a) Euclidean plane}
\hskip -1mm
\subfig{0.4\linewidth}{img/goldberg.pdf}{(b) $\gshort 721$}
\hskip -1mm
\subfig{0.4\linewidth}{img/letters-centered.pdf}{(c) $\ghoo$}
\hskip -1mm
\subfig{0.4\linewidth}{img/letters-sideview.pdf}{(d) not centered}
}
\end{center}
\caption{\label{figlet} Goldberg-Coxeter construction, and $\ghoo$ with labeled vertices, in two perspectives.}
\end{figure}

For our applications it is useful to consider hyperbolic triangulations finer than $\scht{7}$. Such triangulations can be obtained with the {\bf Golberg-Coxeter construction}, which
adds additional vertices of degree 6. 
Consider the $\scht{6}$ triangulation of the plane, and take an equilateral triangle $X$ with one vertex in point $(0,0)$ and another vertex in the point obtained by moving $a$ steps in a
straight line, turning 60 degrees right, and moving $b$ steps more (in Figure \ref{figlet}a, $(a,b)=(2,1)$). 
The triangulation $\gp{a}{b} T$ is obtained from the triangulation $T$ by replacing each of its triangles with a copy
of $X$ (Figure \ref{figlet}b). Regular triangulations are a special case where $a=1, b=0$. For short, we denote the triangulation $\gp{a}{b} \scht{q}$ with $\gqab$.
Figure \ref{figtile}d shows the triangulation $\ghoo$.

Let $v_0$ be a vertex in a hyperbolic triangulation $G$ of the form $\gqab$. 
We denote the set of vertices of $G$ by $V(G)$.
For $v,w \in V(G)$, let $\dist(v,w)$ be the length of the shortest path from $v$ to $w$.
Below we list the properties of our triangulations which are the most important to us.
These properties hold for all hyperbolic triangulations of form $\gqab$;
see Appendix \ref{ommited} for the proof of Properties \ref{prop_canonical}, \ref{prop_expgrow} and \ref{prop_shortcut} for $\gqab$.

\begin{property}[rings]
\label{prop_rings}
The set of vertices in distance $k$ from $v_0$ is a cycle.
\end{property}

We will call this cycle $k$-th {\bf ring}, $R_k(G)$. 
We assume that all the rings $R_k(G)$ are oriented clockwise around $v_0$. Thus, the
$i$-th successor of $v$, denoted $v+i$, is the vertex obtained by starting from $v$
and going $i$ vertices on the cycle. The $i$-th predecessor of $v$, denoted $v-i$,
is obtained by going $i$ vertices backwards on the cycle. A {\it segment} is the
set $S = \{v, v+1, \ldots, v+k\} \subsetneq R_k(G)$ for some $v \in V$ and $k \geq 0$;
$v$ is called the leftmost element of $S$, and $v+k$ is called the rightmost element
of $S$. By $[v,w]$ we denote the segment such that $v$ is its leftmost element, 
and $w$ is its rightmost element. 
For $v,w \in R_k(G)$, let $w-v$ be the smallest $i \geq 0$ such that $w = v+i$.
We also denote $\dist_0(v) = \dist(v,v_0)$. By $B_k(G)$ we denote the $k$-th ball
(neighborhood of $v_0$), i.e.,~$B_k(G) = \bigcup_{i=0,\ldots,k} R_k(G) = \{v \in V | \dist(v,v_0) \leq k\}$.

\begin{property}[parents and children]
\label{prop_pch}
Every vertex (except the root $v_0$) has at most two parents and at least two children. 
\end{property}

We use tree-like terminology for connecting the rings. A vertex $w$ is a {\bf parent} of
$v$ if there is an edge from $v$ to $w$ 
and $\dist_0(v)=\dist_0(w)+1$; in this case, $v$ is a {\bf child}
of $w$. Let $P(v)$ be the set of parents of $v \in R_k(G)$; it forms a segment of 
$R_{k-1}(G)$, and its leftmost and rightmost elements are 
respectively called the {\bf left parent}
$p_L(v)$ and the {\bf right parent} $p_R(v)$. The set of children
$C(v)$, leftmost child $c_L(v)$ and rightmost child $c_R(v)$ are defined analogously.
By $p_L^d$, $P^d$, etc. we denote the function iterated $d$ times, e.g., 
$P^d(v)$ is the set of $d$-th ancestors of $v$, and $p_L^d(v)$ is the leftmost one.

Figure \ref{figlet}cd depicts the triangulation $\ghoo$ with named vertices. Both
pictures use the Poincar\'e disk model and show the same vertices, but the left picture is
centered roughly at $v_0$ (labeled with $A$ in the picture), and the right picture is centered at a different
location in the hyperbolic plane. Points drawn close to the boundary of the 
Poincar\'e disk are further away from each other than they appear -- for example,
vertices $T$ and $U$ appear very close in the left picture, yet in fact all
the edges are roughly of the same length (in fact, there are two lengths --
the distance between two vertices of
degree 6 is slightly different than the distance between a vertex of degree 6 and a vertex
of degree 7).

Vertices $X$, $Y$, and $Z$ are the children of $T$; its siblings are $S$ and $U$,
and its parents are $O$ and $P$  (Fig. \ref{figlet}cd). The values of $P^k([Y])$ for consecutive values
of $k$, i.e., the ancestor segments of $Y$,
 are: $[Y]$, $[T]$, $[O,P]$, $[L,M]$, $[I,K]$, $[F,H]$, $[D,E]$, $[B,C]$, $[A]$.
Vertex $W$ has just a single ancestor on each level: $R$, $N$, $K$, $H$, $E$, $C$, $A$.
Vertex $V$ has the following ancestor segments: $[Q,R]$, $[M,N]$, $[J,K]$, $[G,H]$, $[D,E]$, 
$[B,C]$, $[A]$. Note the tree-like nature of our graph: $[D,E]$ is the segment of 
ancestors for both $V$ and $Y$, and $[O,P]$ and $[Q,R]$ are already adjacent.
This tree-like nature will be useful in our algorithms.

\begin{property}[canonical shortest paths]
\label{prop_canonical}
Let $v, w \in V(G)$, and $\dist(v,w) = d$. Then 
$v \in P^d(w)$ or $w \in P^d(v)$ or 
there exist $a, b, c \in \bbN$ such that
$p_R^a(v) + b = p_L^c(w)$ or
$p_R^a(w) + b = p_L^c(v)$, where $a+b+c = d$. 

\end{property}

In other words, the shortest path between any pair of two vertices $(v,w)$ can always be
obtained by going some number of steps toward $v_0$, moving along the ring, and
going back away from $v_0$. The cases where one of the vertices is an ancestor of
the other one had to be listed separatedly because it is possible that $|P^a(v)| > 2$
for $a > 1$, thus $w$ might be neither the leftmost nor the rightmost ancestor.
Such a situation happens in $\ghoo$ for the pair of vertices labeled
$(J,O)$ in Figure \ref{figlet}, even though $|P^a(v)| \leq 3$
always holds.

In the following, we denote the set of finite words over an alphabet $T$ by $T^*$.

\begin{property}[regular generation]
\label{prop_rgen}
There exists a finite 
set of types $T$, a function $c: T \ra T^*$, and an assignment $t: V(G) \ra T$ 
of types to vertices, such that for each $v \in V(G)$, the sequence of types of
all children of $v$ from left to right except the rightmost child is given by $c(t(v))$.
\end{property}

The rightmost child of $v$ is also the leftmost child of $v+1$, so we do not include
its type in $c(t(v))$ to avoid redundancy. Our function $c: T \ra T^*$ can be uniquely
extended to a homomorphism $T^* \ra T^*$, which we also denote with $c$, in the following way:
$c(t_1\ldots t_k) = c(t_1) \ldots c(t_k)$. By induction, the sequence of types of non-rightmost vertices
in $C^k(v)$ is given by $c^k(t(v))$.

\begin{figure}[h]
\centering
\shortonly{
\subfig{0.2\linewidth}{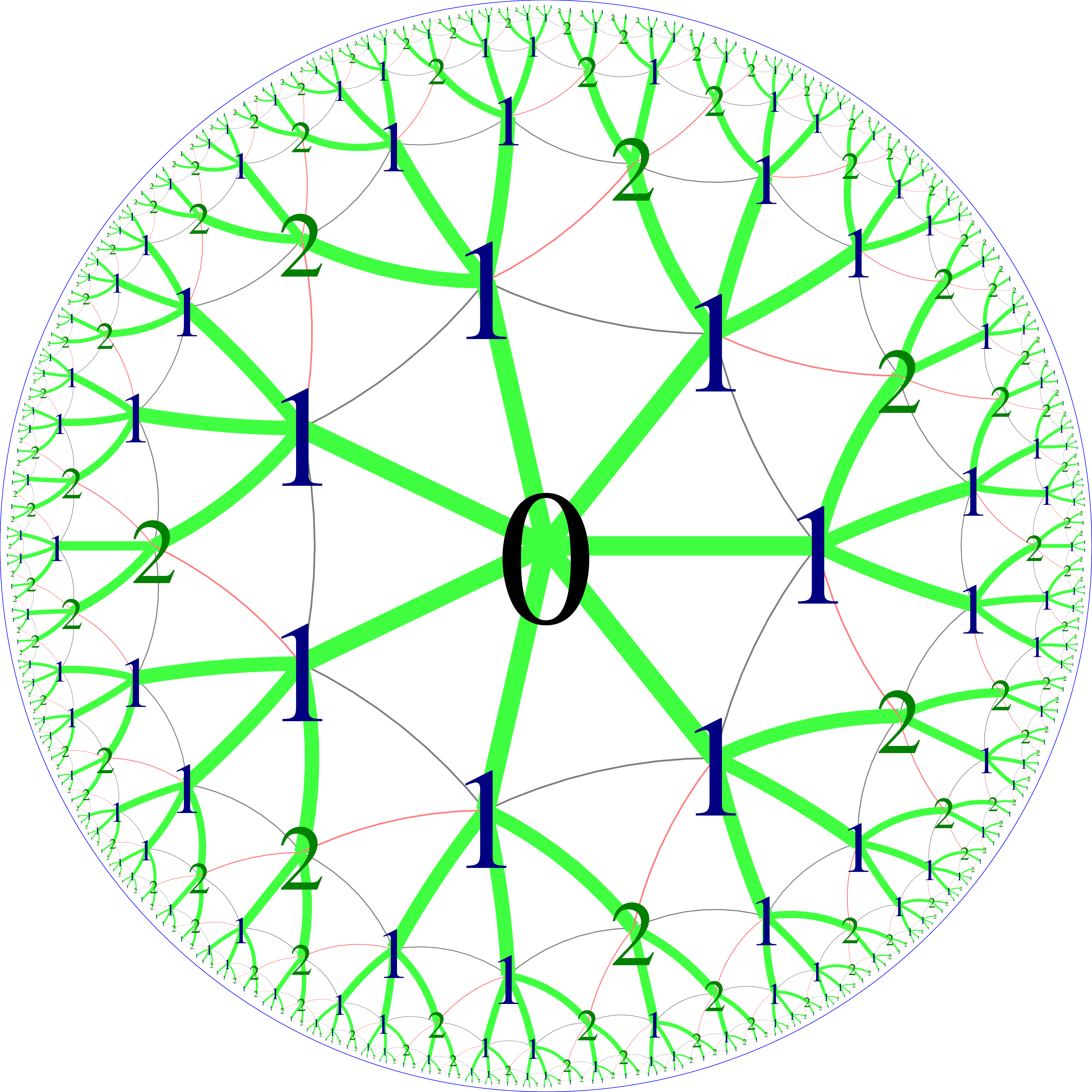}{(a) $\ghoz$}
\hskip -3mm
\subfig{0.2\linewidth}{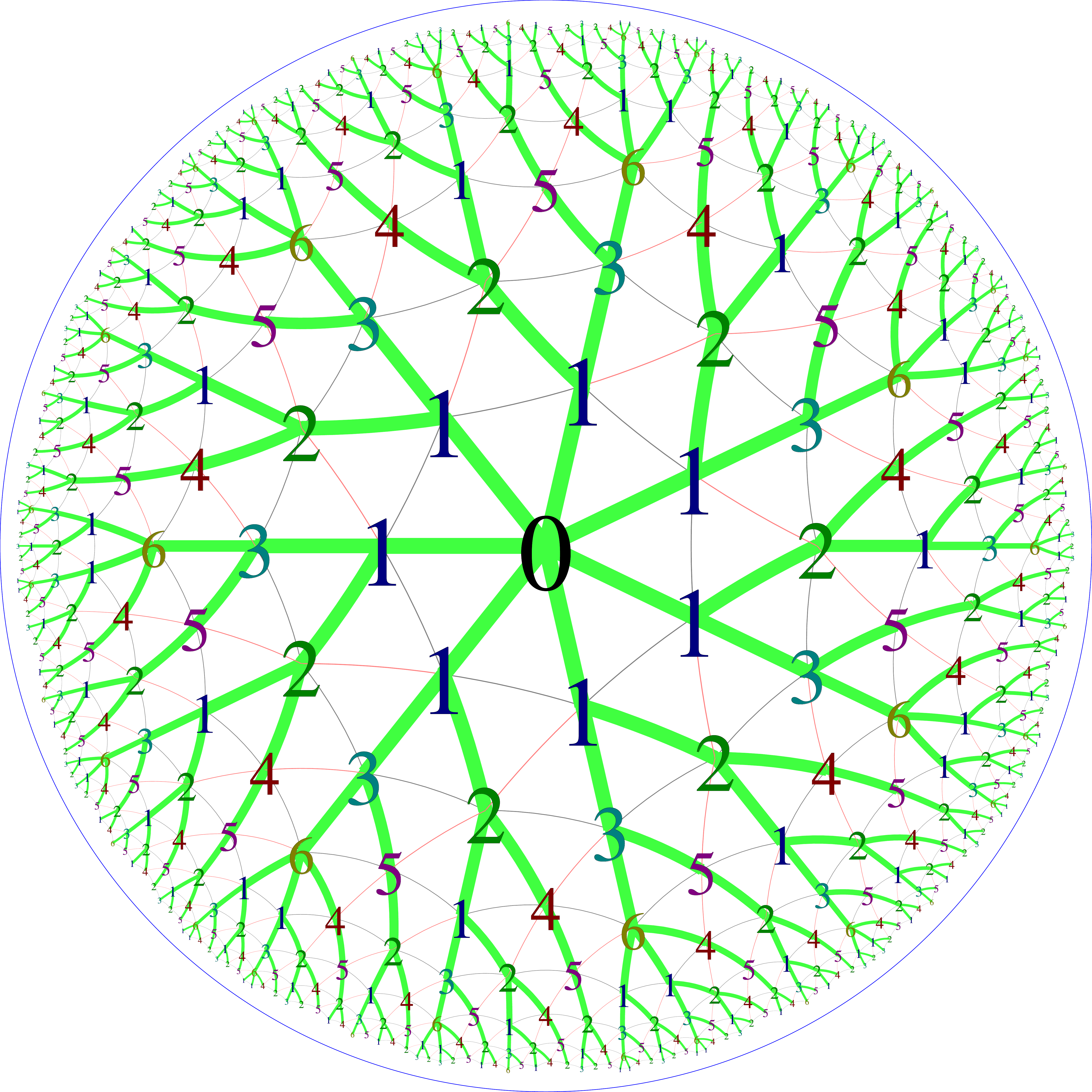}{(b) $\ghoo$}
\hskip -3mm
\subfig{0.2\linewidth}{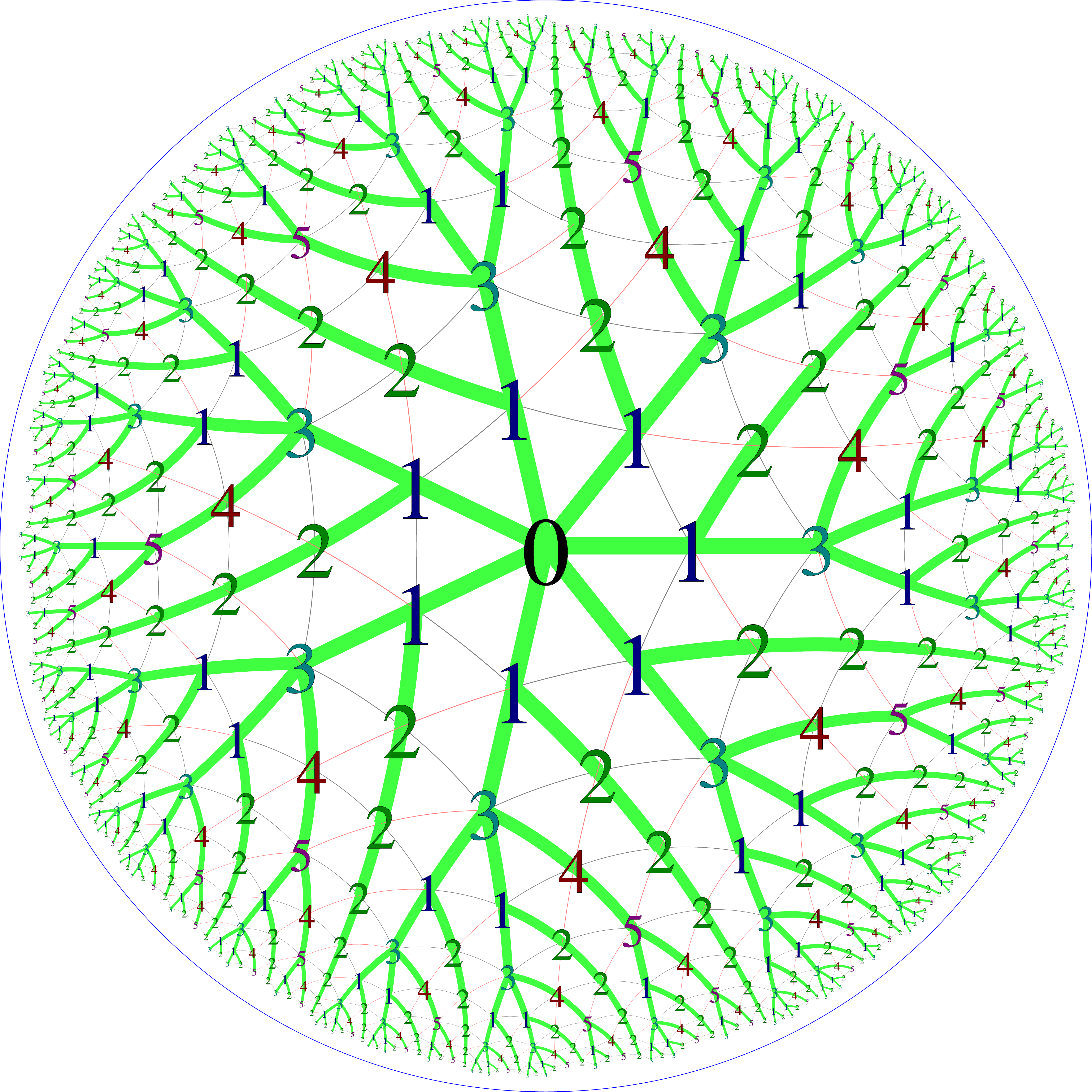}{(c) $\gshort 720$}
\hskip -3mm
\subfig{0.2\linewidth}{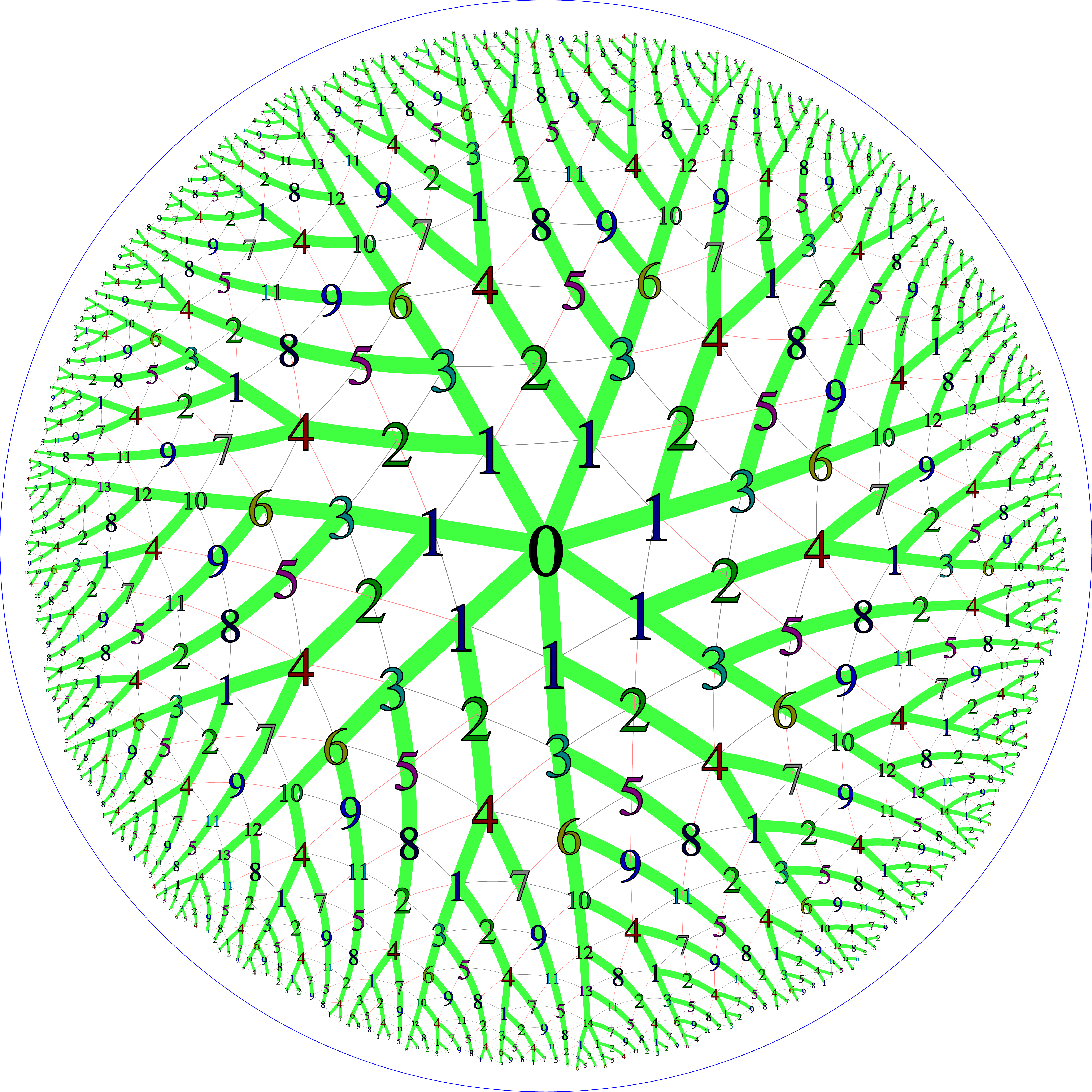}{(d) $\gshort 721$}
\hskip -3mm
\subfig{0.2\linewidth}{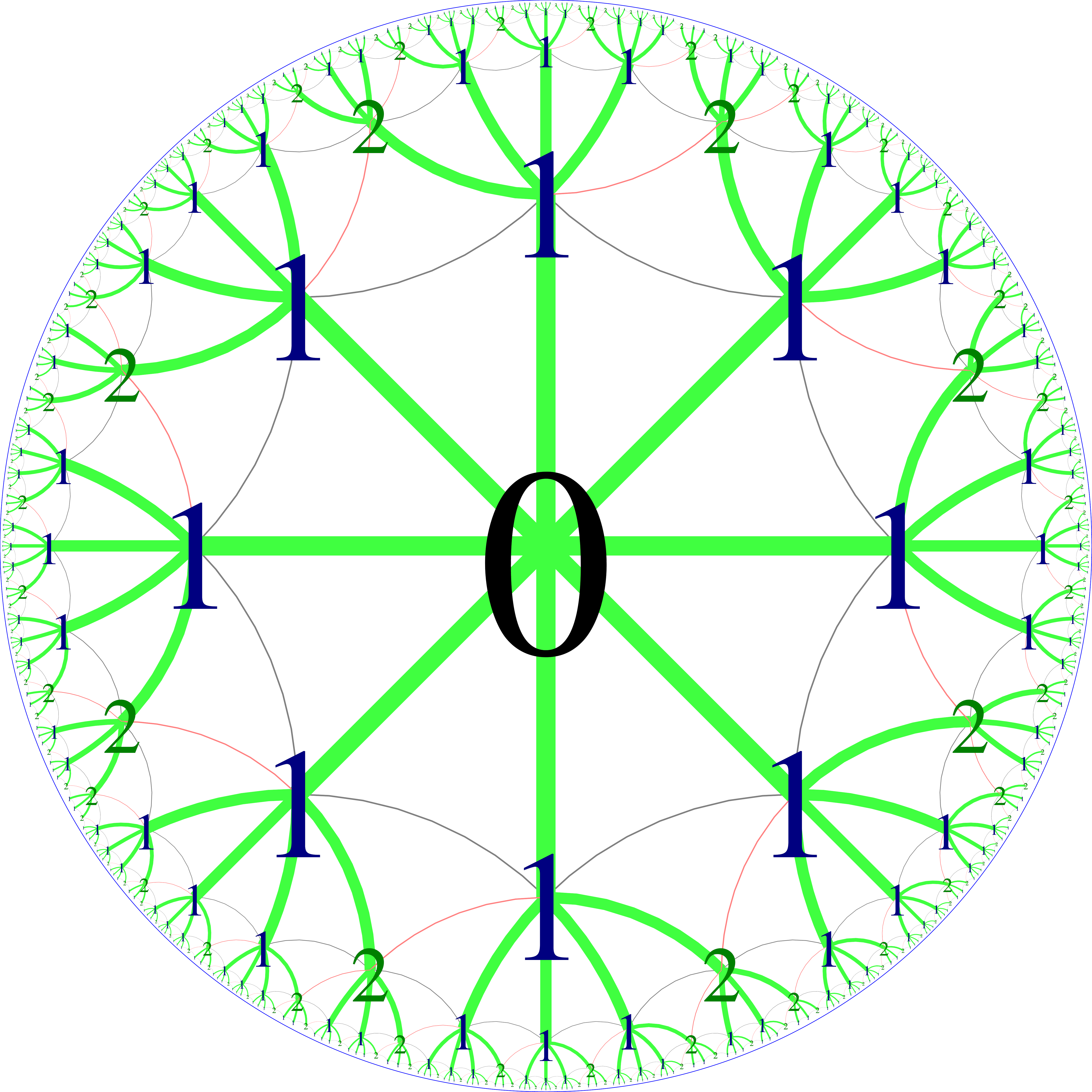}{(e) $\gooz$}
}
\longonly{
\subfig{0.3\linewidth}{tes/710.pdf}{(a) $\ghoz$}
\hskip -3mm
\subfig{0.3\linewidth}{tes/810.pdf}{(e) $\gooz$}
\hskip -3mm
\subfig{0.3\linewidth}{tes/711.pdf}{(b) $\ghoo$}
\hskip -3mm
\subfig{0.45\linewidth}{tes/720.pdf}{(c) $\gshort 720$}
\hskip -3mm
\subfig{0.45\linewidth}{tes/721.pdf}{(d) $\gshort 721$}
}
\caption{Type assignments on RGHTs.\label{typergth}}
\end{figure}

For regular triangulations $\scht{q}$, the set of types is $T=\{0,1,2\}$, and
the types correspond to the number of parents (Figure \ref{typergth}ab). The root has type 0 and has $q$ children
of type $1$, thus $c(0)=1^q$. For a vertex with $t=1,2$ parents, the leftmost child has
type 2 (two parents), and other non-rightmost children all have type 1. Thus,
we have $c(t)=21^{q-4-t}$. Such constructions for $\sch{3}{q}$ and $\sch{4}{q}$ grids have
been previously studied by Margenstern \cite{margenstern2013small,margenstern_pentagrid,margenstern_heptagrid}.

For $\gp{1}{1}$ triangulations there are 7 types, because 
we also need to specify the degree of vertex $v$ as well as the orientation (the degree of the first child)
(Figure \ref{typergth}c).
For Goldberg-Coxeter tesselations in general we need to identify the position of $v$ in the
triangle $X$ used in the Goldberg-Coxeter construction (Figure \ref{typergth}de).

\begin{property}[exponential growth]
\label{prop_expgrow}
There exists a constant $\gamma(G)$ such that, for every vertex $v$,
$|C^{k}(v)| = \Theta(\gamma(G)^k)$.
\end{property}

\def\tlimit{D(G)}

\begin{property}[tree-likeness]
\label{prop_shortcut}
There exists a constant $\tlimit$ such that, for every $d > \tlimit$ and $x \in V(G)$, the distance from $x$ to $x+d$ is smaller than $d$.
\end{property}

This property gives an upper bound on the value of $b$ in Proposition \ref{prop_canonical}, and thus it
will be crucial in our algorithms computing distances between vertices of $G$.
Note that Euclidean triangulations do not have this property.


Given the canonicity of shortest paths and regular generation, the value of $D(G)$ can be found with 
a simple algorithm. The value of $D(G)$ is very small for the grids most important in our applications: $D(\gshort q10)=2$ and $D(\gshort q11)=3$.
For larger values of $a,b$, obtaining $D(\gqab)$ theoretically is challenging. 
We have verified experimentally for $a,b \leq 15$ that $D(\gqab) = 2a+b$.

\begin{definition}
A {\bf regularly generated hyperbolic triangulation} (RGHT) is a triangulation which satisfies all the properties
listed above.
\end{definition}

\section{Segment tree graphs}\label{sec:stg}

If $G$ is a regularly generated hyperbolic triangulation, Properties \ref{prop_canonical} and 
\ref{prop_shortcut} yield a simple algorithm for computing the distance between two vertices 
$v_1, v_2 \in V(G)$ \cite{dhrgex}. For $d = \min\{\delta_0(v_1), \delta_0(v_2)\}, \ldots, 0$, we compute
the segments $s_1 = P^{\delta_0(v_1)-d}(v_1)$ and $s_2 = P^{\delta_0(v_2)-d}(v_2)$, and see if their
distance on the $d$-th ring is at most $D(G)$; if yes, $\delta(v_1,v_2) = (\delta_0(v_1) -d) + (\delta_0(v_2)-d)
+ \delta(s_1,s_2)$.

For example, let's compute the distance between $W$ and $Y$ in Figure \ref{figlet} (d). We have $\delta_0(W)=7$ and $\delta_0(Y)=8$. For $d=7$, we
get two segments $[W]$ and $P(Y)=[T]$, which are still too far. For $d=6$ we have two segments 
$P(W)=R$ and $P(T)=[O,P]$, which are in distance $2 \leq D(G)$, thus $\delta(W,Y) = (\delta_0(W)-d) + (\delta_0(Y)-d) + \delta(s_1,s_2) = 
1 + 2 + 2 = 5$.

This algorithm runs in time $\delta(v_1,v_2)$. Let $\calS \supseteq V$ be
the set of all segments of form $P^k(v)$ for some $v \in V$; from Proposition \ref{prop_shortcut} we know that all segments of this form are
short. While $G$ is not a tree itself, ($\calS$,P) is a tree which provides sufficient information
to compute distances.

Segment tree graphs are a generalization of this construction. We abstract from the definition of a segment in Section
\ref{sec:htgrid} and treat our segments as abstract objects. This lets our method work with not only
RGHTs, but also other graphs, such as higher-dimensional hyperbolic tessellations and other Gromov hyperbolic graphs, 
such as Cayley graphs of hyperbolic groups.

In our data structure, the set of vertices $V$ is 
embedded as a subset of the set of segments $\calS$, which forms a tree. We can use the structure of the tree $\calS$
to compute the distance between $v_1, v_2 \in V$ in a way which is a generalization of the above: we compute $P^k(v_1),
P^l(v_2) \in \calS$ until we find two segments $s_1, s_2$ which are on the same level and ``close'', and then 
$\delta(v_1, v_2)) = k + l + \delta(s_1,s_2)$. The relation $N \subseteq \calS \times \calS$ is the set of pairs
of ``close'' segments, and thus, knowing $N$ and $\delta_N$, which is $\delta$ restricted to $N$, lets us efficiently compute the distance
between any pair of vertices or segments. As will be explained later (Theorem \ref{thm:stg}),
our structure also allows to answer more complex queries efficiently.

\begin{definition} \label{def:stg}
A {\bf $(m_N,m_d)$-bounded segment tree graph} is a tuple $(\calS, V, v_0, P, \delta_0, N, \delta_N, \delta)$ such that:
\begin{itemize}
\item $V \subseteq \calS$ represents the set of vertices (we call the elements of $\calS$ segments),
\item $v_0 \in V$ is the root vertex,
\item $P: {\calS \setminus v_0} \ra \calS$ is the parent segment function,
\item $\delta_0: \calS \ra \bbN$ is the depth function: for $s \in \calS$, $P^{\delta_0(s)} = v_0$,
\item $N \subseteq \bigcup_{d\in\bbN} (P^{-d}(v_0))^2$ is a symmetric and reflexive relation such that
if $\{s_1,s_2\} \in N$ and $s_1 \neq v_0$, then $\{P(s_1), P(s_2)\} \in N$
(where $P^d$ is $P$ iterated $d$ times, and
$P^{-d}(s)$ for $s\in \calS$ denotes $\{s': P^d(s')=s\}$),
\item $\delta_N: N \ra \bbN$ is the near distance function,
\item $|N(s)|$ is bounded by $m_N$, and $\delta_N$ is bounded by $m_d$.
\end{itemize}

We say that a segment tree graph:

\begin{itemize}
\item {\bf realizes} a graph $G$ if $V_\calS = V_G$ and $\delta_\calS(v_1,v_2) = \delta_G(v_1,v_2)$,
where $\delta_\calS(s_1, s_2) = k + l + \delta_N(P^k(s_1), P^l(s_2))$, where $k, l \in \bbN$ are the 
smallest such that $(P^k(s_1), P^l(s_2)) \in N$ (note: since the relation $N$ is only defined
for segments at the same depth, we have $k-l = \delta_0(s_1) - \delta_0(s_2)$, thus a smallest
pair is well defined);

\item is {\bf efficient} iff all the operations $s\in V$, $P(s)$, $P^{-1}(s)$, $\delta_0(s)$,
$N(s)$, $\delta_N(s_1,s_2)$ can be performed in amortized time $O(1)$.

\item is {\bf regular} if and only if there is a 
type function $T:\bigcup_{d\in\bbN} (P^{-d}(v_0))^* \ra \bbN$ such that $T(s_1, \ldots, s_c) \in \{1, \ldots, H_c\}$, and
whenever $T(s_1, \ldots, s_c) = T(s'_1, \ldots, s'_c)$:

\begin{itemize}
\item if $(s_i, s_j) \in N$, then $(s'_i, s'_j) \in N$ and $\delta_N(s_i, s_j) = \delta_N(s'_i, s'_j)$,
\item for every $\phi: \{1..b\} \ra \{1..c\}$ there is a bijection $r$ between $C=\prod_{i=\{1..b\}} P^{-1}(s_{\phi(i)})$
and $C' = \prod_{i=\{1..b\}} P^{-1}(s'_{\phi(i)})$ such that $T(t) = T(r(t))$ for $t \in C$.
\end{itemize}

\item is {\bf efficient regular} if it is efficient and regular, and the type function $T$ can also be computed in time O(1) for fixed $c$.
\end{itemize}
\end{definition}

\begin{theorem}\label{rght:stg}
If $G$ is a RGHT, then there exists a efficient regular segment tree graph $\calS$ which realizes $G$.
(We treat $D(G)$ as a fixed constant.)
\end{theorem}

\begin{proof}
Fix a RGHT $G$. We can represent $V$, the set of vertices of $G$ using handles (pointers); 
every vertex knows pointers to its left and right sibling, left and right parent, and its children that
have been already computed. This structure is built on top of the underlying tree, where every
vertex has pointers to its (right) parent and (non-rightmost) children; in particular, when a new 
vertex is generated, so are all its right ancestors. The sibling edges are used for efficient
computation -- when the vertex is queried for the given neighbor for the first time, the required 
vertex is generated or found in the tree (based on the tree structure), but the result is cached
for future use.

As mentioned above, when generating a new vertex, some of its ancestors sometimes also need to be generated. 
We can use the accounting method to show that vertices and edges can be generated in amortized time $O(1)$. 
To every vertex $v$
which is not yet connected to its right sibling we associate $k$ credits, where $k>0$ is the smallest
number such that $v+k+1$ has the same right $i$-th ancestor as $v+k$. Similarly, to every vertex which is not yet connected
to its left sibling we associate $k$ credits, where $k>0$ is such that $v-k-1$ has the same right $i$-th ancestor as $v-k$.
From Proposition \ref{prop_shortcut} we know that there exists $i$ such that both values of $k$ are less than $D(G)$, so
generating the new vertex together with the credits itself will cost $O(1)$, and the cost of generating and linking
its new ancestors will be covered by the credits in the ancestors of $v$.

Our $\calS$ will be the set of all segments $[s_L,s_R]$ which are of form $P^k(v)$. 
Proposition \ref{prop_shortcut} gives an $O(1)$ bound on $s_R-s_L$.
Given our representation of $V$, it is straightforward to implement $v_0$, $s \in V$, $P(s)$,
$P^{-1}(s)$, $\delta_0(s)$ in amortized time O(1). Taking all segments of length at most
$D(G)$ ensures that, for every $v \in V$, all segments of form $P^k(v)$ appear in $\calS$.

Our relation $N$ will consist of pairs of segments $(s,t)$ which are
either intersecting, or $t_L = s_R + x$ for $1 \leq x \leq D(G)$, or $t_R = s_L + x$
for $1 \leq x \leq D(G)$. Set $\delta_N(s,t)$ to be the smallest distance $\delta_G(v_s,v_t)$
for $v_s \in s$, $v_t \in t$. Since $\delta_N(s,t) \leq D(G)$ and $|s|, |t| \leq D(G)$,
$\delta_N(s,t)$ can be computed in O(1), e.g., by checking all the possibilities.

From Propositions \ref{prop_canonical} and \ref{prop_shortcut} we can see that, for $v_1, v_2 \in G$,
$\delta(v_1,v_2)$ from Definition \ref{def:stg} equals $\delta_G(v_1,v_2)$. 

Regularity follows from the regular generation of $G$. Let $(s_1, \ldots, s_c) \in \calS^c$. 
In our type, we record the types of all vertices in $s_1, \ldots, s_c$ (say, in the order
from left to right), as well as the distances $\delta_N(s_i,s_j)$. For a fixed $c$ the number of
possible types is bounded.
\end{proof}

We can now state our main result. Fix an efficient segment tree graph $\calS$ and a connected
colored graph $C$. We can construct
a structure which represents a coloring of $B_R$ that can be changed dynamically.
For the current coloring can query about the number of embeddings $m:V_C \ra B_R$ where,
for every $w \in V_C$, $m(w)$ is of color $k_C(w)$, and the distances between $m(w)$ 
are prescribed by the function $d: V_C \cup E_C \ra [0..2R]$, which is the argument of the query. 
We in fact allow more general colorings, that could be seen as functions $\val_k: B_R \ra \bbR$ for every $k \in K$. 
In a typical coloring every vertex has at most one color, i.e., $\val_k(v)=1$ for at most one $k$ and $\val_k(v)=0$ for all other colors. 
In this more general setting, vertices can be given fractional or multiple colors, and
each embedding counts as $\prod_{w \in {V_C}} \val_{k_C(w)}$.

\begin{theorem}\label{thm:stg}
Fix an efficient segment tree graph $\calS$, a finite set of colors $K$,
and a connected colored graph $C=(V_C,E_C,k_C:V_C\ra K)$. 
Let $R \in \bbN$. 
Let $B_R = \bigcup_{d \leq R} P^{-d}(v_0) \cap V_\calS$ be the ball of radius $R$ in $V_\calS$.
Then there exists a data structure with the following operations:

\begin{itemize} 
\item $\InitCounter$, which initializes $\val_k: B_R \ra \bbR$ to 0 for every $k \in K$,
\item $\Add(k, v, x)$, which adds $x$ to $\val_k(v)$, where $v \in B_R$ and $k \in K$,
\item $\Count(d)$, which for $d: V_C \cup E_C \ra [0..2R]$ returns the following:

\[ \sum_{m:V_C \ra B_R} \left( \prod_{w \in {V_C}} \val_{k_C(w)}(m(w)) \right) \gamma_d(m), \]

where $\gamma_d(m)$ is 1 if and only if $\delta(m(a), m(b)) = d(a, b)$ for every $(a,b) \in E_C$, and
$\delta_0(m(w)) = d(w)$ for every $w \in V_C$, and 0 otherwise.
\end{itemize}

Such $\Count$ and $\InitCounter$ can be implemented in $O(1)$, and $\Add$ can be implemented in $O(R^{|E_C|+|V_C|})$.
\end{theorem}


\begin{proof}
For $D$, a connected subgraph of $C$, let $u: V_D \ra \calS$ be such that every $u(w)$ has the same depth $\delta_0(u)$, 
and for every $(a,b) \in E_D$ we have $(u_a, u_b) \in N$. Denote the set of all descendant segments of $s$ by $P^{-*}(s)$.
For $d: V_D \cup E_D \ra [0..2R]$ define: ($[\phi]$ is 1 if $\phi$ is true, 0 otherwise)

\begin{equation} \label{defcount}
c_D(u,d) = \sum_{m:V_D \ra B_R} \left(\prod_{w \in {V_D}} [m(w) \in P^{-*}(u(w))] \val_{k_C(w)}(m(w)) \right) \gamma_d(v).
\end{equation}

Our algorithm keeps the value $c_D(u,d)$ for 
every $D$, $u$ and $d$ such that there exists a $v$ which is a descendant of some $u(w)$ ($w \in V_D$) and for which an $\Add$ operation
has been performed (otherwise we know that $c_D(u,d)$ is 0). After every $\Add$ operation, our
algorithm will update the changed values of $c_D(u,d)$. Since $\Count(d) = c_C((v_0, \ldots, v_0), d)$, this lets us perform the
$\Count$ operation in $O(1)$.

In a non-dynamic setting,
we could compute $c_D$ using recursion with memoization, by examining all the possible embeddings $m$ in formula (\ref{defcount}). 
Every $m(w)$ can either equal $u(w)$, or can be in $P^{-*}(u'(w))$ for some $u'(w) \in P^{-1}(u(w))$. We examine all the possible
subsets $W \subseteq V_D$, and for each $w \in W$, we examine all the possible $u'(w) \in P^{-1}(u(w))$; for $w \in W \setminus V_E$
we take $u'(w) = u(w)$. Let $c_{D,W}(u,u',d)$ be defined as in (\ref{defcount}), but where $m(w)$ is additionally restricted
to $P^{-*}(u'(w))$ for $m(w) \in W$, and to $u(w)$ for $m(w) \notin W$. The value of $c_D(u,d)$ is the sum of obtained values
$c_{D,W}(u,u',d)$ over all choices of $W$ and $u'$.

We will show how to compute $c_{D,W}(u,u',d)$. Let $E_W$ be the subset of $E_D$ of edges $(a,b)$ such that $a,b \in W$ and
$(u(a), u(b)) \in N$. We need to count $m:V_D \ra B_R$ such that: 
\begin{itemize}
\item $\delta(m(a), m(b)) = d(a, b)$ equals $d(m(a)) + d(m(b)) - 2\delta_0(u) + \delta_N(u(a), u(b))$ for every $(a,b) \in E_D \setminus E_W$,
\item $\delta_0(m(a))$ equals $d(w)$ for every $w \in V_D$,
\item $\delta(m(a), m(b))$ equals $d(a, b)$ for every $(a,b) \in E_W$.
\end{itemize}

If the first condition is not satisfied, we do not need to count anything. Otherwise, if the graph $(W, E_W)$ is connected
we get the requested value by calling $c_{(W,E_W)}(u',d)$ recursively. If the graph $(W, E_W)$ is not connected, we split it into connected
components $(W_1, E_{W_1})$, \ldots, $(W_n, E_{W_n})$ and get the requested value by multiplying
$c_{(W_i,E_{W_i})}(u' \restto{W_i}, d \restto{W_i \cup E_{W_i}})$ for all $i \in 1..n$, and $\val_{k_i}(u(w))$ for all $w \in D \setminus W$.

In the dynamic setting, we recompute the changed $c_D$ values after every $\Add$ operation.
When we call $\Add(k, v, x)$, we need to recompute $c_D(u, d)$ for every
subgraph $D$ of $C$, for every
$u$ such that $v$ is a descendant of some $u(w)$,
and for every $d$ such that $d(w)=\delta_0(v)$ for some $w \in V_D$.
There are only at most $R+1$ such descendants, $O(R^{|E_C|+|V_C|-1})$ possible choices
of $d$, and since $u$ is a connected subgraph of $(\sum_d P^{-d}(v_0), N)$, $O(|V_C|! m_N^{|V_C|-1})$ possible choices of other segments in $u$.
Therefore, we can update all the necessary values of $c_D(u,d)$ in time $O(R^{|V_C|+|E_C|})$.
\end{proof}

\begin{theorem} \label{thm:stgd}
Additionally, if our $\calS$ is regular, we can have an alternative initialization operation 
$\InitCounter(k_0)$, where $k_0 \in K$. This operation initializes $\val_{k_0}$ to 1 for every $v \in B_R$,
and can be performed in time $O(R^{|V_C|+|E_C|+2})$.
\end{theorem}

\begin{proof}
The proof is the same as for Theorem \ref{thm:stg}, except that we no longer have $c_D(u,d)=0$ if $\Add$ operation
has never been performed for any descendant of $u(w)$. However, we know that 
$c_D(u,d) = c_D(u',d)$ if $u$ and $u'$ are in the same distance from $v_0$, and
$T(u) = T(u')$. Therefore, we can compute $c_D(u,d)$ just once for every type and in every distance from 0 to $R$, 
thus the initialization can be performed in time $O(R^{|V_C|+|E_C|+2})$. 
\end{proof}

{\bf Remark.} In Theorem \ref{thm:stg}, $\Count(d)$ requests specific distances for
every edge in $E_C$, and for technical reasons also distance from $v_0$ for every vertex in $V_C$.
However, in our applications, we usually do not want to know the complete information.
For example, we may not care about distances from $v_0$, or we may only want to count the edge distances which satisfy
a specific relation, e.g., $d(m(w_1),m(w_2)) < d(m(w_2),m(w_3))$. It is straightforward to
adjust the proof of Theorem \ref{thm:stg} to counting such embeddings, possibly obtaining
a smaller exponent in the time complexity of the $\Add$ operation.

\section{Generalization to Gromov hyperbolic graphs}\label{sec:other}

A {\it geodesic} in the graph $(V,E)$ is a path $\gamma = (\gamma_0, \ldots, \gamma_d)$ which is a shortest path from
$\gamma_0$ to $\gamma_d$, i.e., $d = \dist(\gamma_0, \gamma_d)$. A {\it geodesic triangle} is
a triple of paths $(\gamma^1, \gamma^2, \gamma^3)$ such that the endpoints of $\gamma^1$ are
some $v_1, v_2 \in V$, the endpoints of $\gamma^2$ are $v_2, v_3 \in V$, and the endpoints of $\gamma_3$
are $v_3, v_1 \in V$. For $A \subseteq V$, the $d$-neighborhood of $A$, denoted $N^d(A)$, is the set of all vertices of $V$ 
in distance at most $d$ from $A$. We say that $G$ is $\delta$-hyperbolic (in the sense of Gromov) if every geodesic triangle
is $\delta$-thin, i.e., for every geodesic triangle $(\gamma^1, \gamma^2, \gamma^3)$ we have $\gamma_3 \subseteq N^\delta(\gamma_1 \cup \gamma_2)$.
For example, trees are 0-thin. For RGHTs, according to Proposition \ref{prop_canonical} the shortest path between $v$ and $w$ have canonical shapes
consisting of a part of the shortest path from $v$ to $v_0$, a part of the shortest path from $w$ to $v_0$, and a middle segment;
Proposition \ref{prop_shortcut} limits the length of this middle segment, and thus the parameter $\delta$, to $O(D(G))$.

\begin{theorem}\label{seggromov}
Let $(V,E)$ be a $\delta$-hyperbolic graph of bounded degree, and $v_0 \in V$. Then there exists a bounded segment tree graph which realizes $(V,E)$.
If $\delta(v,v_0)$ and the set of neighbors of $v$ can be computed in $O(1)$, then this segment tree graph is efficient.
\end{theorem}

\begin{figure}[h!]
\centering  
\subfig{0.4\linewidth}{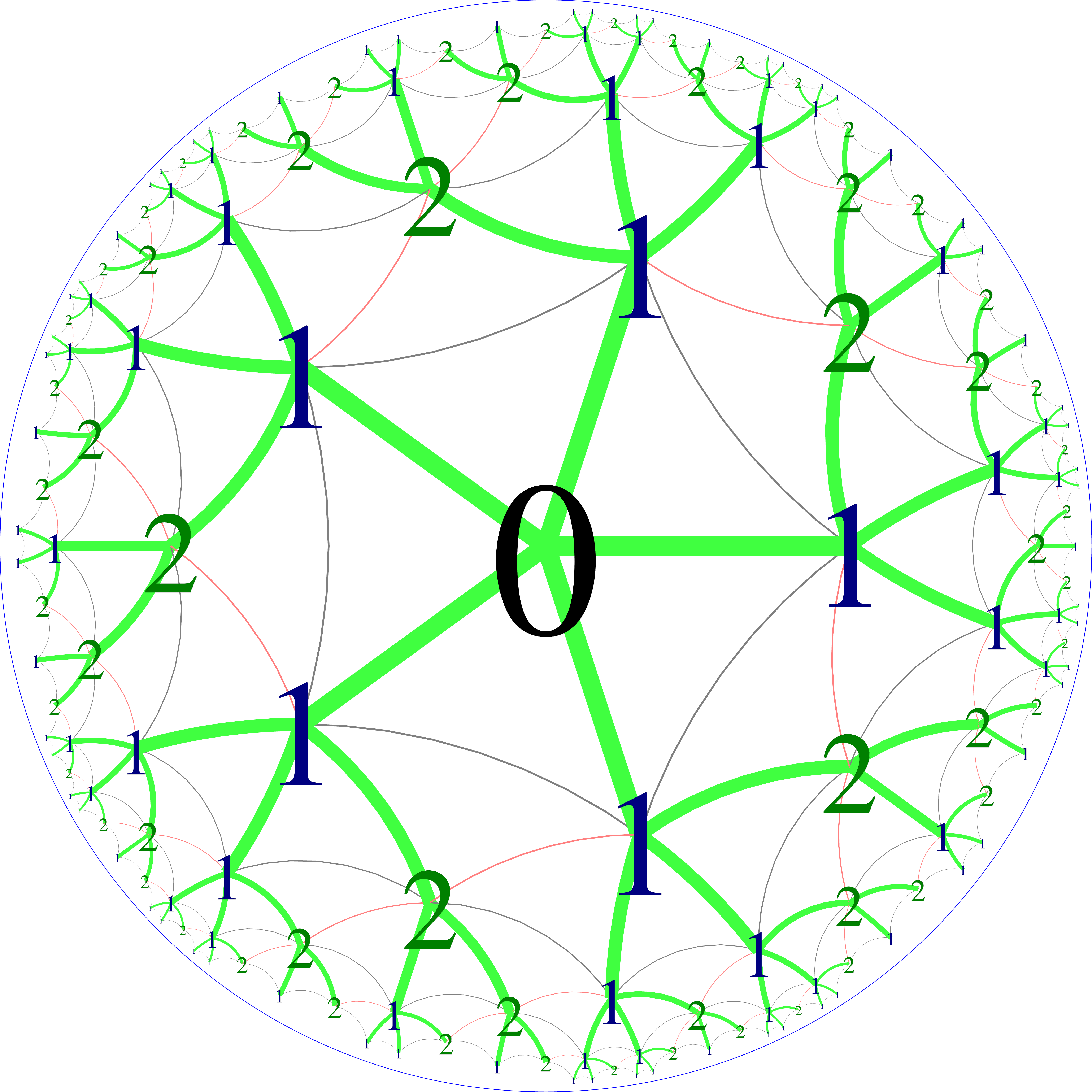}{(a) $G_{4,5}$}
\subfig{0.4\linewidth}{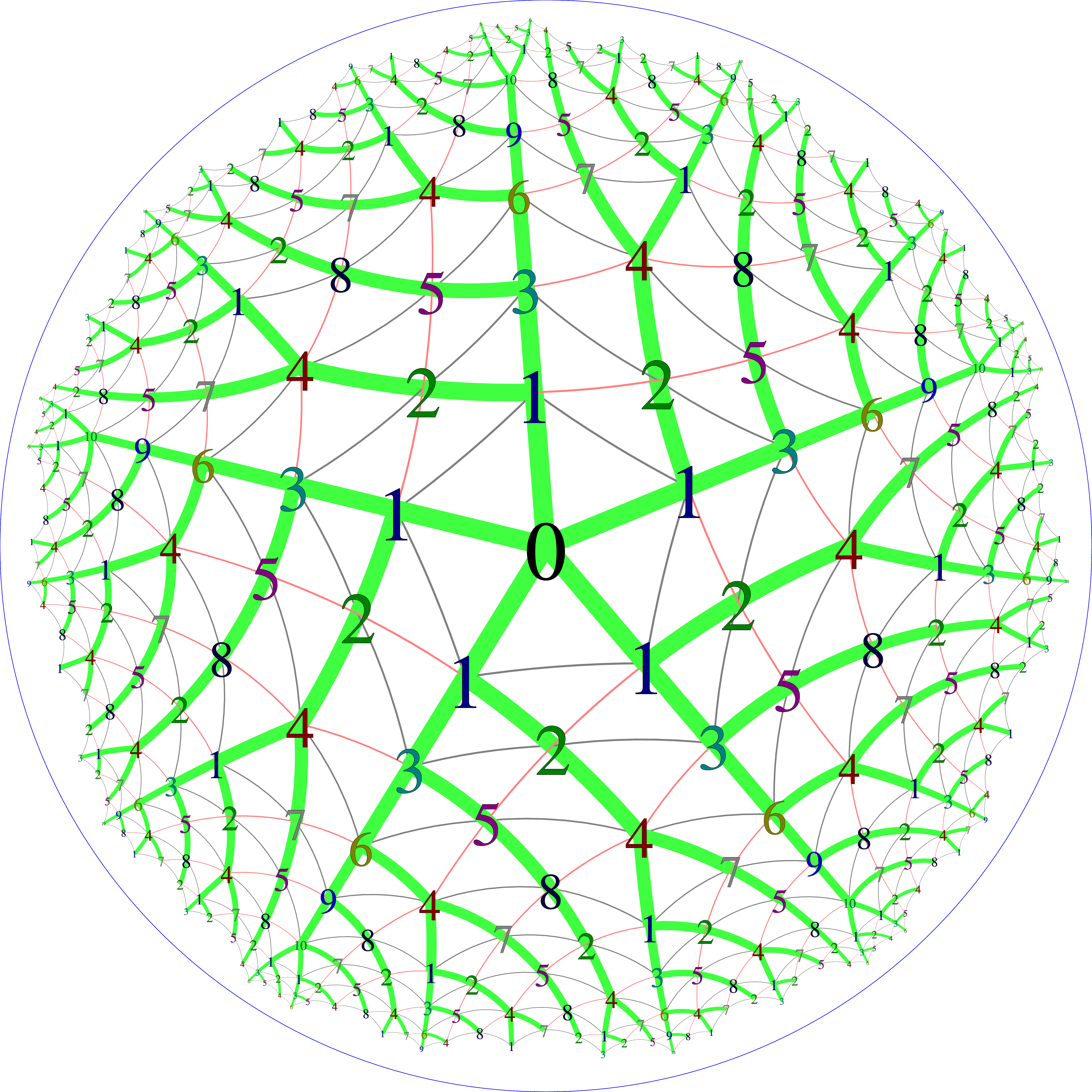}{(b) $GC_{2,1} G_{4,5}$}
\subfig{0.4\linewidth}{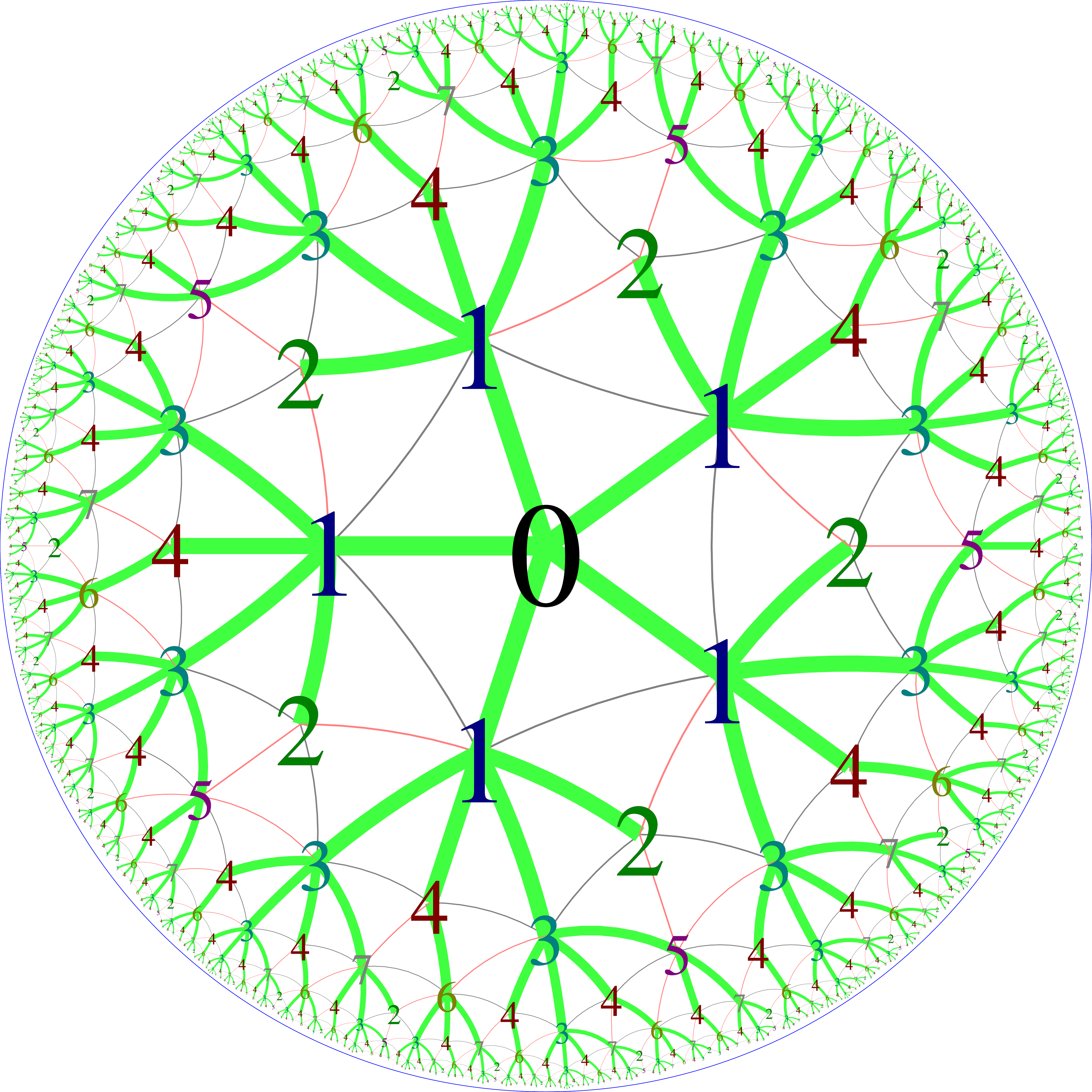}{(c) $V8.8.5$}
\subfig{0.4\linewidth}{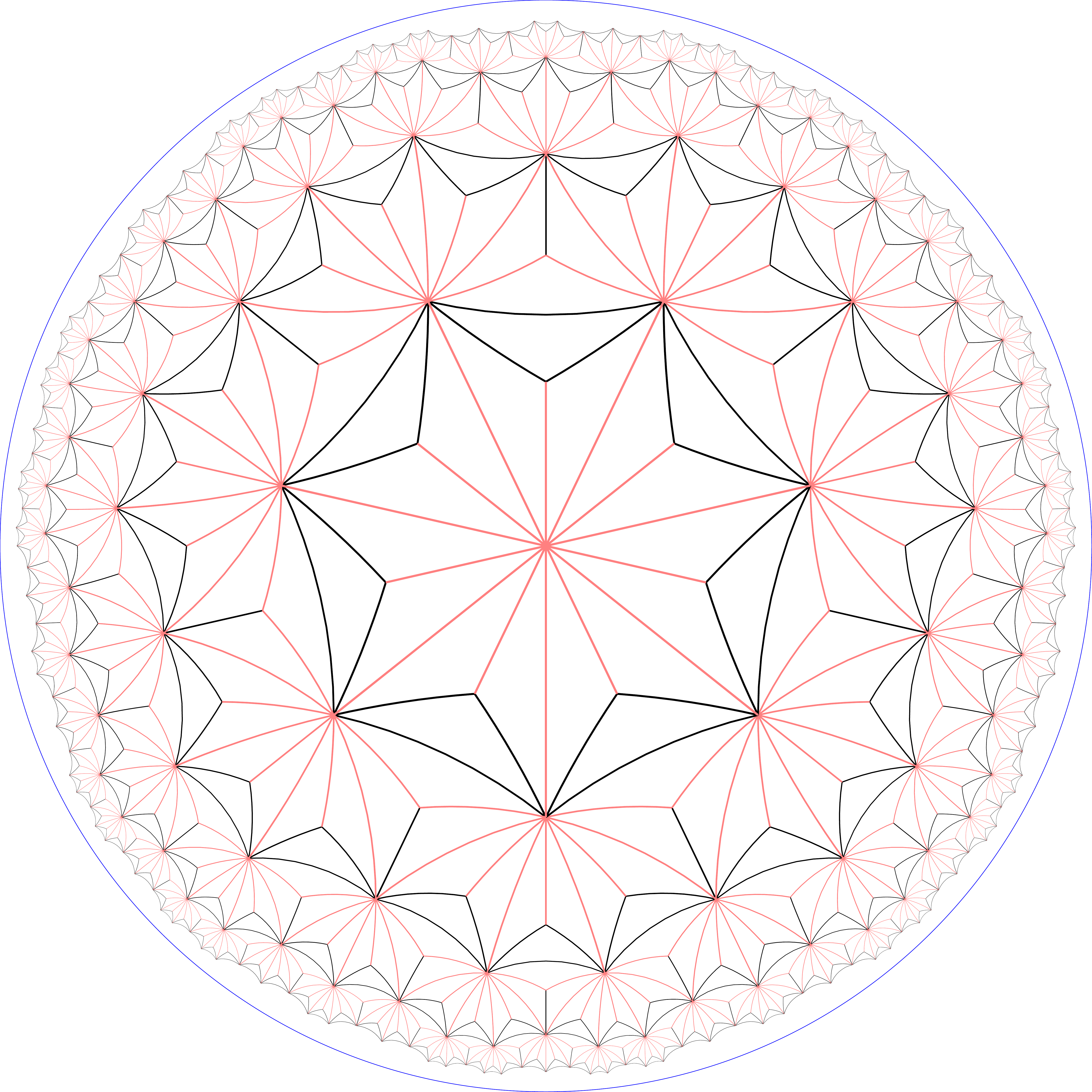}{(d) $V14.14.3$}
\caption{Square grids and Catalan tesselations.}\label{typeother}
\end{figure}

The proof of Theorem \ref{seggromov} is technical, and can be found in Appendix \ref{ommited}.
The segments constructed in the general proof have unpractical, irregular structure compared
to the BSTG constructed for RGHTs in Theorem \ref{thm:stg}, so it is useful to find
tessellations for which simpler, regular constructions work. 
In Section \ref{sec:htgrid} we have explored triangulations of form $G_{qab}$. We can also 
explore quadrangulations, i.e., $\gp{a}{b} \schq{q}$ for $q \geq 5$ (Goldberg-Coxeter construction for quadrangulations is
defined similarly as for triangulations -- we use the square grid). The major difference here is that the rings $R_k(G)$ are disconnected rather than cycles.
However, this only makes our algorithms simpler: the canonical shortest paths (Proposition \ref{prop_canonical}) no longer have to go
across the ring, i.e., $b$ always equals 0.
However, Proposition \ref{prop_canonical} fails for $\sch{p}{q}$ where $p > 4$. There are face-transitive (Catalan) triangulations
where the sets of vertices in distance $d$ from $v_0$ do not form rings; for example, the triangulation with face configuration V8.8.5
(Figure \ref{typeother}c) has vertices with three parents; this causes the tree-like distance
property (Proposition \ref{prop_shortcut})
to fail (consider a vertex $v$ with 3 parents and the shortest path from the leftmost parent of $v$ to $v+1$). If we
split every face of $\scht{7}$ into three isosceles triangles, we obtain the triangulation with face configuration V14.14.3 (Figure \ref{typeother}d),
where the sets $R_k(G)$ are no longer cycles (vertices repeat on them), causing the regular generation to fail.
Such cases are less relevant for our applications, because they give less accurate approximations of the hyperbolic plane
(square tilings already give worse results in our applications).

\begin{figure}[h!]
\centering  
\subfig{0.4\linewidth}{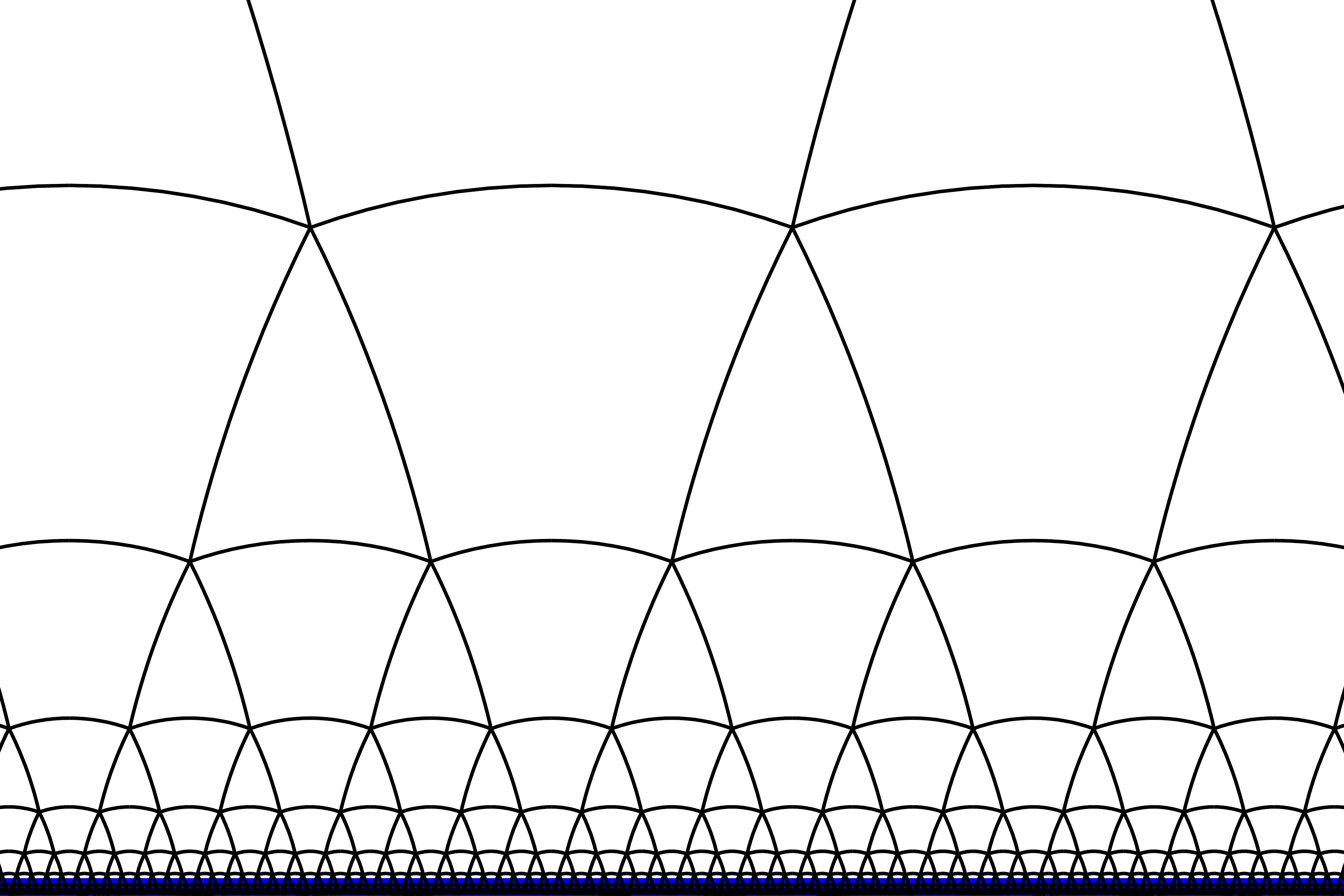}{(a) binary grid}
\subfig{0.4\linewidth}{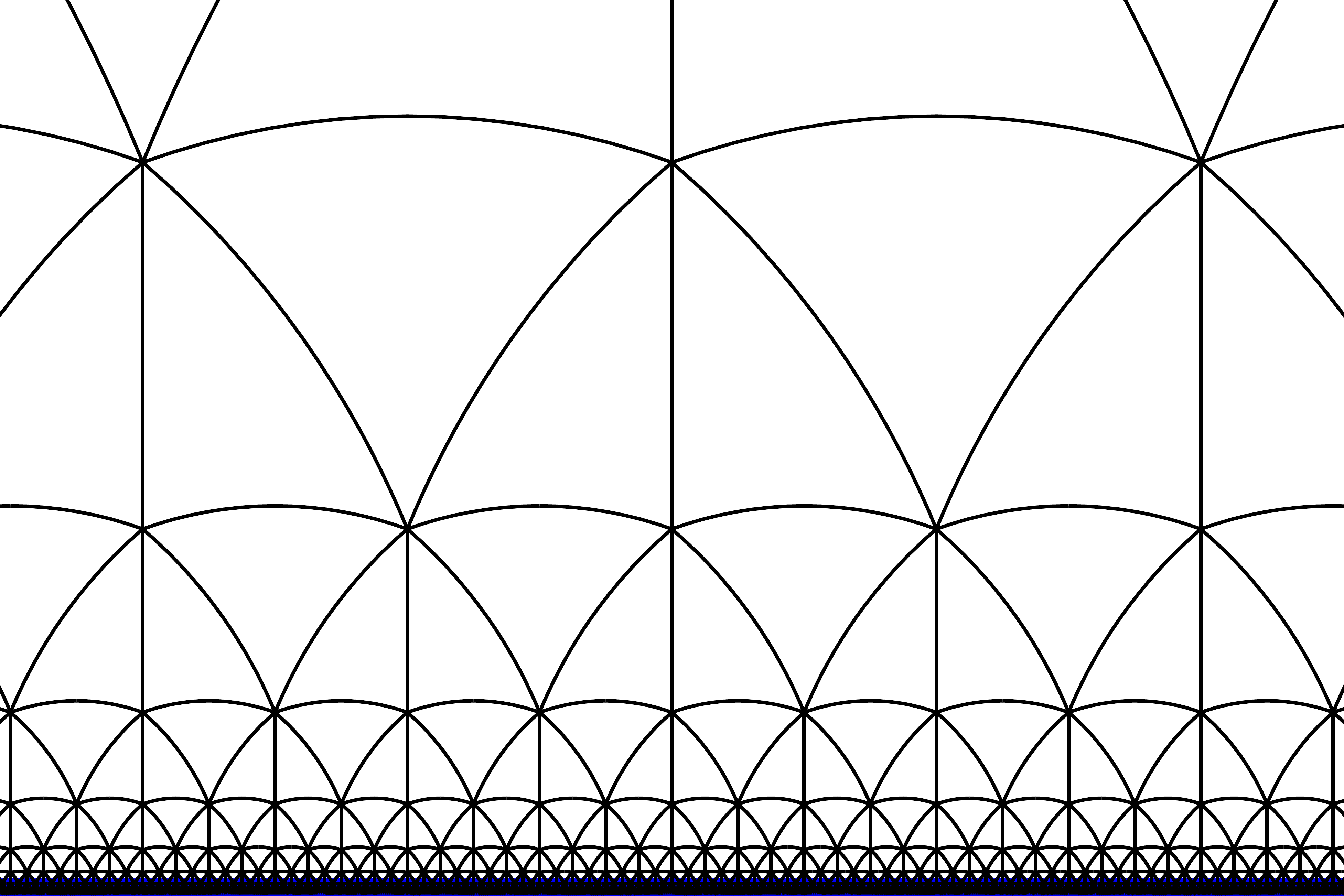}{(b) variant}
\caption{The binary grid, and a variant. \label{binaries}}
\end{figure}

The binary grid (dual of the binary tiling \cite{boroczky}) is shown in Figure \ref{binaries}a.
Figure \ref{binaries} uses the Poincar\'e upper half-plane model, where the scale is smaller
closer to the bottom line. 
It is a very simple tessellation of the hyperbolic plane which yield a very simple BSTG structure.
It also generalizes to higher dimensions.

\begin{definition}
The $d$-dimensional binary grid is the graph $G_d = (V,E)$ where $V = \mathbb{Z}^d$.
Let $e_i \in \mathbb{Z}^d$ have 1 in $i$-th coordinate and 0 in other coordinates.
Every vertex $v$ is connected with an edge to $v+e_i$ and $v-e_i$ for $i=1, \ldots, d-1$,
as well as its children $(2v_1+c_1, 2v_2+c_2, \ldots, 2v_{d-1}+c_{d-1}, v_d+1)$, where
$c_i \in \{0,1\}$.
\end{definition}

The set of all descendants of $0$ in $G_d$ forms a segment tree graph. (Considering only
descendants of 0 makes our graph a bit asymmetrical; there are many ways to improve this,
which we do not list here for brevity.) In these segment tree graphs,
$\calS=V$, and $(v_1, v_2) \in N$ iff $\delta_0(v_1) = \delta_0(v_2)$ and $v_1-v_2$ has all
coordinates between -4 and 4; for $(v_1, v_2) \in N$, $\delta_N(v_1, v_2)$ is the distance between
$v_1$ and $v_2$. It is easy to show that $\delta$ as in Definition \ref{def:stg} equals 
the distance function in $G_d$.

We can also define a variant binary grid, where the offsets $c_i$  are allowed to be -1, 0, or 1.
This corresponds to a slightly different construction shown in Figure \ref{binaries}b. Again, this is 
a segment tree graph, but now segments may be 1 or 2 vertices wide in every coordinate. The relation
$N$ can be defined similarly as above. Both kinds of variant binary tilings yield efficient regular
segment tree graphs (there is only one type of a vertex).

Higher-dimensional segment tree graphs show that, while our algorithms are based on tree-likeness,
they are not restricted to graphs of bounded treewidth. In fact, $p^{-k}(0)$ in $G_3$ is 
the $2^k \times 2^k$ square grid, which has treewidth $2^k$; and the graph $\cup_{i=0}^k p^{-i}(0)$
has $O(2^k)$ vertices. Note that hyperbolic distances between these grid points are 
approximately logarithms of Euclidean distances, so our methods can be also used in Euclidean spaces
when we are only interested in approximate distances, up to a multiplicative factor.

\section{Applications}\label{sec:apps}

\paragraph*{Hyperbolic Random Graph model} Our results have potential applications in social network analysis.
Take the hyperbolic plane $\bbH^2$ with a designated central point $h_0$. 
The Hyperbolic Random Graph (HRG) model creates a random graph $H = (V,E)$ as follows:

\begin{itemize}
\item Each vertex $v \in V = \{1, \ldots, n\}$ is randomly assigned a point $\mu(v) \in \bbH^2$,
by randomly choosing the distance from $\mu(v)$ to $h_0$ (according to a fixed distribution) 
and direction (according to the uniform distribution).

\item Every pair of vertices $v_1, v_2 \in V$ is connected with an edge with probability
$p(\dist(\mu(v_1), \mu(v_2)))$, where $\dist(\mu(v_1), \mu(v_2))$ is the hyperbolic distance from $\mu(v_1)$
to $\mu(v_2)$, and $p$ is some function, e.g., $p(x) = 1 / (1 + \exp(Tx+R))$ where $R$ and $T$ are parameters of the model. Closer
points are more likely to be connected.
\end{itemize}

For correctly chosen parameters, the HRG model generates graphs with properties
(such as degree distribution and clustering coefficient) similar to that of real-world scale-free
networks. However, using the continuous hyperbolic plane $\bbH^2$ may raise problems.
The first problem is that navigating in continuous hyperbolic geometry may be difficult to understand. The second problem is that
all coordinate-based representations of $\bbH^d$ are prone to precision issues because of the
exponential growth \cite{tobias_alenex}: the area of a hyperbolic circle of radius $R$ is of the order of $\exp(R(d-1))$,
hence any representation using $k$ real numbers represented with $l$ bits will collapse some points
into a single one if $R / \log 2 > kl / (d-1)$. This issue can be completely avoided by using discrete
tessellations of the hyperbolic plane; geometry of such a discrete tessellation is very similar to
that of the underlying continuous hyperbolic geometry \cite{hyperrogue}, and we do not lose precision,
because distances between vertices in the HRG model are typically large relative to the edge length
of the tessellation. Another benefit is that we get to use the techniques from this paper to easily obtain
efficient algorithms for the relevant computations. The experimental results will 
be discussed in detail in another paper \cite{dhrgex}
; here we present it just as an example area of application.

\paragraph*{Choosing the parameters}
In the continuous model, we can use calculus to compute the expected degree distribution and clustering
coefficient of the obtained graph. The clustering coefficient is the probability that vertices $a$ and $c$ are
connected with an edge, under a condition that vertices $a$ and $b$ are connected, as well as $b$ and $c$.
In the discrete variant of the HRG model (DHRG), this is more
difficult; however, we can compute the expected values for the given parameters by using Theorem \ref{thm:stgd}.
See Appendix \ref{sec:dhrg_app} for more details.

\paragraph*{Generating HRGs} The brute-force method of generating HRGs works by considering every pair of vertices,
and connecting them according to the computed probability. This is inefficient, 
and there have been many papers devoted to generating HRGs efficiently.
The original paper \cite{papa} used an $O(n^3)$ algorithm. Efficient algorithms have been found
for generating HRGs in time $O(n)$ \cite{gengraph} and for MLE embedding real world scale-free networks
into the hyperbolic plane  in time ${\tilde O}(n)$ \cite{tobias}, which was a major improvement over 
previous algorithms \cite{hypermap,vonlooz}, and recently in $O(n)$ \cite{hypefficient}.
Our discrete
model lets us generate DHRGs efficiently and easily. We use a graph $C$ consisting of a single edge with
endpoints of colors $k_1$ and $k_2$. We generate all the vertices of our network, and give them color $k_1$.
Then, for every vertex $v \in V$, we color $\mu(v)$ with color $k_2$; this lets us find out how many
vertices are in distance $d$ from $v$, which lets us to batch process them, and generate a DHRG more efficiently
(see Appendix \ref{sec:dhrg_app}).

\paragraph*{Embedding HRGs} Another important problem is MLE embedding of real-life networks. We are given a network
$H = (V,E)$, and we want to find an embedding $m: V \ra \bbH^2$ which maximizes the log-likelihood, which is defined as
$\sum_{v_1,v_2 \in V} \log_{(v_1,v_2)\in E} p(d(\mu(v_1), \mu(v_2)))$, where $\log_\phi p = \log p$ if $\phi$ is true,
and $\log(1-p)$ if $\phi$ is false. (Intuitively, the log-likelihood is the probability of obtaining such a graph randomly 
using the HRG method.) Embedding is a difficult problem, as even computing the log-likelihood via brute force requires
$O(n^2)$ time. In \cite{tobias} an algorithm is given to compute approximate log-likelihood, and to embed networks, in time $O(n)$. 
In the DHRG model, we can use Theorem \ref{thm:stg} to compute the number of pairs of vertices in every distance;
this not only gives a $O(nR^2)$ algorithm to compute the log-likelihood, but also we can re-compute the log-likelihood
after moving a vertex of degree $a$ in time $O(R^2+aR)$. (Since the graph $C$ is very simple in this case, and we do not care about distances from
$v_0$, we get a better exponent than the general one from Theorem  \ref{thm:stg}.)
Our experiments show that, despite using a discrete approximation, 
we get a better estimate of continuous log-likelihood than the method from \cite{tobias}, and furthermore, we can improve an
embedding by locally moving vertices in order to improve the log-likelihood -- this is not only more efficient than the $O(n^2)$
method given in \cite{tobias}, but also turns out to produce higher quality embeddings when remapped back to the continuous
hyperbolic plane. \cite{dhrgex}

\paragraph*{Pseudo-betweenness} A major issue in social network analysis is to find the important nodes in the network. 
This is done using {\it centrality measures} -- functions $f: V \ra \bbR$ which say how important $v \in V$ is. One example
of a centrality measure is the {\it betweenness centrality} $b$. $b(v)$ is defined as $\sum_{v_1, v_2 \in V} b'(v_1,v,v_2)$,
where $b'(v_1,v,v_2)$ is the fraction of shortest paths from $v_1$ to $v_2$ which go through $v$. Unfortunately, computing 
betweenness is computationally expensive (Johnson algorithm $O(|V|\times |E|)$).
The DHRG model 
lets us define pseudo-betweenness using the same formula, but where $b'(v_1,v,v_2) = \gamma^{\delta(v_1,v)+\delta(v,v_2)-\delta(v_1,v_2)}$;
for $\gamma=0$ we get 1 if $v$ is directly on the shortest path from $v_1$ to $v_2$, and 0 otherwise; for a larger value of $\gamma < 1$,
we we also give weight if $v$ is not directly to the path, but close to it. In both cases, Theorem \ref{thm:stg} over a triangle with
two vertices of color $k_2$ and one vertex of color $k_1$ lets us to compute the pseudo-betweenness of every vertex in time $O(nR^{O(1)})$.
This can be done by first coloring all vertices
with color $k_2$, and then for every $v$, we compute its psuedo-betweenness by temporarily coloring $v$ with $k_1$. Again, the
exponent is smaller than one computed in Theorem \ref{thm:stg}, since we are not interested in the distances from $v_0$,
and the special form of our formula lets us compute the result more efficiently than in the general case. Experimental evaluation of this is a subject of a
future paper.

\longonly{
\paragraph*{Machine learning.} Recently hyperbolic embeddings have found application in machine learning. The idea is very similar to DHRG model,
although embeddings are evaluated using other metrics; just like with DHRGs, mass computing of distances lets us to efficiently evaluate and
improve embeddings while avoiding numerical errors.}

\longonly{\paragraph*{General algorithms.} Graphs with structure typical to hyperbolic geometry appear in computer science; examples include 
skip lists which essentially use randomly generated hyperbolic graphs to construct an efficient dictionary, as well as Fenwick trees, quadtrees and octrees which
are essentially based on the binary tiling and its higher dimensional variants. 
\nonblind{The paper \cite{planarspectra},
where all the basic constructions are essentially hyperbolic graphs. }Understanding hyperbolic graphs may lead to new discoveries in computer science.}

\paragraph*{Other applications} Hyperbolic tessellations are used in data visualization \cite{hrviz}, unsupervised learning \cite{ritter99,ontrup},
and video games \cite{hyperrogue,hypminesweeper}. Efficient algorithms for generating such tessellations and computing distances in them are indispensable in these applications.


\newpage

\appendix
\section{Omitted proofs}\label{ommited}

\begin{proof}[Proof of Property \ref{prop_canonical} for $\gqab$]

Let $v, w \in V(G)$ for a triangulation $G$ satisfying the previous properties.
Let $(v=v_0, v_1, v_2, \ldots, v_d=w)$ be a path from $v_0=v$ to
$v_d=w$ of length $d$. We will show that a path from $v$ to $w$
exists which is of the form given in Proposition \ref{prop_canonical}
and is not longer than $d$.

In case if $v \in P^d(w)$ or $w \in P^d(v)$, the hypothesis trivially
holds, so assume this is not the case.

\def\ras#1{\stackrel{#1}{\ra}}

\def\ca{{v_i}}
\def\cb{{v_{i+1}}}
\def\cc{{v_{i+2}}}

Each edge from $\ca$ to $\cb$ on the path is one of the following types:
right parent, left parent, right sibling, left sibling,
right child (inverse of left parent, i.e., any non-leftmost child), left child
(inverse of right parent, i.e., any non-rightmost child).
We denote the cases as respectively $\ca \ras{RP} \cb$,
$\ca \ras{LP} \cb$, $\ca \ras{RS} \cb$, $\ca \ras{LS} \cb$, 
$\ca \ras{RC} \cb$, $\ca \ras{LC} \cb$. 
We use the symbols $x,y$
if we do not care about the sides.

If $\ca \ras{xC} \cb \ras{yP} \cc$, then we can make the path shorter
($\ca$ and $\cc$ are both parents of $\cb$ and thus, from
Proposition \ref{prop_rings}, they must be the
same or adjacent).

If $\ca \ras{xS} \cb \ras{yP} \cc$, then let $u$ be such that $\ca \ras{yP} u$.
Either $u=\cc$ or $u$ is adjacent to $\cc$, so we can replace this situation
with $\ca \ras{yP} \cc$ or $\ca \ras{yP} u \ras{zS} \cc$, without making the
path longer. The case $\ca \ras{yC} \cb \ras{xS} \cc$ is symmetric.

Therefore, all the $xP$ edges must be before all the $xS$ edges, which must
be before all the $xC$ edges. Furthermore, clearly all the $xS$ edges must
go in the same direction -- two adjacent edges moving in opposite directions
cancel each other. 

We will now show that all the edges have to go in the same direction (right 
or left). This direction will be called $m \in \{L,R\}$. There are three cases:
\begin{itemize}
\item there are $xS$ edges -- if they do not all go in the same direction,
then two adjacent ones moving in the opposite directions cancel each other,
so we can get a shorter path by removing them. Otherwise, let $m$ be the
common direction.
\item there are no $xS$ edges, and the vertex between $xP$ edges and $xC$
edges is the root -- in this
case, we get from $v$ to the root using $a$ parent edges, and then from 
the root to $w$ using $c$ child edges. If we replace the first $a$ edges
with right parent edges, we still get to $v_0$; symmetrically, we replace
the last $c$ edges with right child edges.
\item there are no $xS$ edges, and the vertex between $xP$ edges and $xC$
edges is $v_i$ which is not the root -- then, the main direction is $R$
iff $v_{i-1}$ is to the left from $v_{i+1}$ among the children of $v_i$,
and $L$ otherwise.
\end{itemize}

Now, we can assume that all the edges in the $xP$ part go in the same
direction (i.e., they are $mP$
edges). Indeed, if this is not the case, let $m'$ be the opposite of $m$,
and take the last $m'P$ edge: $\ca \ras{m'P} \cb \ra \cc$. The
$\cb \ra \cc$ edge could be a $mP$ edge if $\ca\ra\cb$ is not the last $P$ edge,
or a $mP$ edge if it is the last and sibling edges exist, or a $mC$ edge
otherwise. In all
cases, let $u$ be such that $\ca \ras{mP} u$. By case by case analysis,
we get that $\ca \ra u \ra \cc$ the path is either shorter (i.e., $u=\cc$)
or pushes the $m'P$ edge further the path. Ultimately, we get no $m'P$
edges in the $xP$ part. By symmetry, we also have no $m'C$ edges in the
$mC$ part.

Therefore, our path consists of $a$ $mC$ edges, followed by $b$ $mS$ edges,
followed by $c$ $mP$ edges. This corresponds to the last two cases of
Proposition \ref{prop_canonical} (depending on whether $m$ is $R$ or $L$),
therefore proving it.
\end{proof}

\begin{proof}[Proof of Property \ref{prop_expgrow} for $\gqab$]
If $v \neq v_0$ and $c(t(v)) = t_1 \ldots t_n$, the number of vertices in $C^k(v)-C^k(v+1)$ is given by 
$\sum_{i=1}^n |c^{k-1}(t_i)|$. This gives a linear recursive system of formulas for computing $g_k(t(v)) = |C^k(v)-C^k(v+1)|$; 
by the well-known properties of such systems, $g_k(t(v))$ grows exponentially or polynomially. However, since every vertex in 
$\gqab$ has a descendant
(in a bounded number of generations) of degree $q>6$ which has more than two children, the growth cannot be polynomial.
Let $\gamma = \max_t \lim_{k \ra \infty} \sqrt[k] {g_k(t)}$; the maximum value has to be obtained for $v=v_0$, and also for every $v$ with degree $q$,
since the type of all children of $v_0$ also appears as a child of every $v$ of degree $q$.
In the grids $\gqab$ every vertex $v$ will eventually produce $w \in C^k(v)$ of degree $q$, thus
$|C^{k}(v)| = \Theta(\gamma(G)^k)$ for every $v$. We have $\gamma(\ghoz) \approx 2.6180339$ and $\gamma(\ghoo) \approx 1.72208$.
\end{proof}

\begin{proof}[Proof of Property \ref{prop_shortcut} for $\gqab$]

We will show how to compute $\tlimit$ algorithmically based on the previous properties.
We initialize $D$ (the current lower bound on $\tlimit$) to 0, and call 
the function {\tt find\_sibling\_limit($v_1$, $v_2$)} for every pair of vertices in $R_1(G)$.
That function compute $v_2-v_1$, and check whether it is smaller than
the length of a path which goes through lower rings; if yes, we update $D := \max(D, v_2-v_1)$. 
Then, {\tt find\_sibling\_limit} calls itself recursively for every $(w_1, w_2)$ where $w_1$ which is non-rightmost child of $v_1$
and every $w_2$ which is non-leftmost child of $v_2$.

This ensures that every pair of vertices is checked. Of course, this is infinitely many pairs.
However, recursive descent is not necessary if:
\begin{itemize}
\item (A) there is a vertex $v_3$ in the segment $[v_1+1, v_2-1]$ which produces an extra child in every generation.
In this case, let $w_1$ be a non-rightmost child of $v_1$ and $w_2$ be a non-leftmost child of $v_2$.
Since every vertex in the segment $[v_1, v_2]$ has at least two children, $w_1$ is non-rightmost, $w_2$ is
non-leftmost, and $v_3$ has at least three children, we have $w_2-w_1 \geq v_2-v_1 + 2$; on the other hand,
$\dist(w_1,w_2) \leq \dist(v_1,v_2) + 2$. By similar argument, the same will be true in the further generations.
\item (B) another pair $(v_1, v_2)$ previously considered had the same sequence of types of vertices in $[v_1, v_2]$,
and the same distances from $v_1$ to $v_2-1$ and from $v_1$ to $v_2$ (the results for any pairs of the descendants
of the current pair would be the same as the results for the respective pairs of descendants of the earlier pair).
\end{itemize}

For our hyperbolic grids $\gqab$, the vertices which produce an extra child in every generation are the ones
of degree $q>6$, and ones of degree 6 which have a single parent.
A segment $[v_1,v_2]$ of vertices of degree 6 with two parents each is locally similar to a straight line in Figure
\ref{figlet}a. Therefore, if its length is greater than $O(a+b)$, one of its descendant segments will eventually 
include a marked vertex, corresponding to a a vertex of degree $q$ in $\gqab$; therefore, the recursive descent 
for such segments will terminate, due to rule (A).
On the other hand, there are finitely many types of segments of shorter length. Therefore the
algorithm will terminate due to rule (B). The final value of $D$ equals $\tlimit$.

\end{proof}

\begin{proof}[Proof of Theorem \ref{seggromov}]
Take a graph $G = (V,E)$.
Let $\delta_0(v) = \delta(v,v_0)$. 
For every vertex $v$, fix a geodesic $\gamma(v)$ from $v$ to $v_0$ in such a way that if $v'$ lies on $\gamma(v)$, then $\gamma(v')$ is
a suffix of $\gamma(v)$. In other words, the edges used by all the geodesics $\gamma(v)$ form a breadth-first search tree of
the graph $G$.

Also let $R_d$ be the $d$-th sphere, i.e., the set of all vertices $v$ such that $\delta_0(v)=d$. 

For $v \in V$ and $k < \delta_0(v)$, let $P^k(v) = (A,h,j)$, where:

\begin{itemize}
\item $A = N^\delta(\gamma(v)) \cap \bigcup_{d=0..r} R_{\delta_0(v)-k+d}$, where $r = 2\delta$,
\item $h:A \ra [0..\delta]$, and $h(w)$ is the distance from $\gamma(v)$ to $w$,
\item $j:A \ra [0..\delta]$, where $j(w) = \dist(v, w) - \dist_0(v)$.
\end{itemize}

For $k = \delta_0(v)$ let $P^k(v) = v_0$. Let $\calS = \{ P^k(v): v \in V, k \in [0..\delta_0(v)]\}$. We need to check whether this $\calS$ satisfies the
definition of the segment tree graph. We need to check the following properties:

\begin{itemize}
\item $V \subseteq \calS$, i.e., $\calS$ has elements which correspond to vertices of $G$. The vertex $v_0$ corresponds to $v_0$,
and $v \neq v_0$ corresponds to $P^0(v)$. Note that, if $P^k(v) =  (A,h,j)$, we have $h(w) = j(w) = 0$ for every $w \in A$
which is on $\gamma(v)$; for $k=0$ there will be just one such vertex, for $k>0$ there will be more.

\item $\calS$ forms a tree -- we need to check that $P$ is a well-defined parent function, i.e., if $P^k(v_1) = P^l(v_2) \neq v_0$, 
then also $P^{k+1}(v_1) = P^{l+1}(v_2)$. If $k = \delta_0(v_1)-1$, then also $l = \delta_0(v_2)-1$, and thus $P^{k+1}(v_1) = P^{l+1}(v_2) = v_0$.

Otherwise, let $P^k(v_1) = (A, h, v)$, $P^{k+1}(v_1) = (A', h', v')$, and $P^{l+1}(v_2) = (A'', h'', v'')$.

The difference between $A$ and $A'$ is that $A$ includes vertices in $R_{\delta_0(v)-k+r}$ which are not included in $A'$, 
and $A'$ includes vertices in $R_{\delta_0(v)-k-1}$ which are not included in $A$.
So the only difference between $A'$ and $A''$ could be for some $w \in R_{\delta_0(v)-k-1}$. 
Such $w$ must be in distance $h'(w) \leq \delta$ from some point $w' \in \gamma(v_1)$. That point must be in either in $A$ or on the part of $\gamma(v_1)$ which
is in $N^{\delta_0(v)-k-1}(v_0)$; in both cases, we get that also $w \in A''$ and $h''(w) = h'(w)$.

We also need to show that $j'(w) = j''(w)$. Consider a geodesic $\gamma$ from $w$ to $v$. 
We will show that there exists $i$ such that $\gamma_i \in A$. If $\gamma(w,v) \leq r$, then $v_0$ itself is in $A$.
Otherwise, let $\gamma'$ be the geodesic from $w$ to $w' \in \gamma(v_1)$ of length $h(w)$.
Consider the geodesic triangle ($\gamma, \gamma', \gamma(v))$. From the definition of $\delta$-hyperbolicity we have
that $\gamma_{r+1}$ must be in $N^\delta(\gamma(v))$ or $N^\delta(\gamma')$; the second case is impossible
because $r+1 \geq \delta + h(w)$, so the first case holds, and $\delta_0(\gamma_{r+1})$ must be such that $\gamma_{r+1} \in A$.

If $\gamma_i \in A$, then we have $j(\gamma_i) = j'(w) - i - \dist_0(v)$. The same reasoning holds for $j''(w)$, so $j'(w) = j''(w)$.

\item We need to define $N$ and $\delta_N$. 
Let $\gamma$ be the geodesic from $v$ to $v'$. By the definition of $\delta$-hyperbolicity, Every vertex in $\gamma$ is in $N^\delta(\gamma(v))$
or in $N^\delta(\gamma(v'))$. Take $i$ such that $\gamma_i \in N^\delta(\gamma(v))$ and $\gamma_{i+1} \in N^\delta(\gamma(v'))$. There must be
a $k$ such that $\gamma_i \in A$ were $P^k(v)=(A,h,j)$ and $\gamma_{i+1} \in A'$ where $P^l(v') = (A',h',j')$, where $\delta^0(P^k(v)) = \delta^0(P^l(v'))=d$.

We have $\delta(v,v') = \delta(v,\gamma_i) + 1 + \delta(v',\gamma_{i+1} = j(\gamma_i) + \dist_0(v) + 1 + j'(\gamma_{i+1}) + \dist_0(v') = 
j(\gamma_i) + j'(\gamma_{i+1}) + 1 + k+d + l+d$.

Taking $\delta_N((A,h,j), (A',h',j')) = 2d+1+j(w)+j'(w')) $, where $w\in A$ and $w'\in A$ is chosen so that $\delta_N$ is minimized, yields 
the correct formula $\delta(v,v') = k+l+\delta_N(P^k(v),P^l(v'))$. However, it is possible that to compute $\delta(v,v')$ we should not take the first $k$ and $l$ such
that $A$ and $A'$ are adjacent -- a higher pair of numbers might yield a better distance. Therefore, the segments $S = (A,h,j)$ and $S' = (A',h',j')$ 
are in relation $N$ if $w \in A$ and $w' \in A'$ are adjacent, at least one of $w, w'$ is in $R_{\delta_0(S)}$, and we never get $\delta_N(P^m(S), P^m(S')) + 2m \leq \delta_N(S,S')$ for $m>0$.

Since our $\delta_N$ finds a path which is going through a vertex in $R_{\delta_0(v)-k}$, we have $\delta_N \geq 0$, and since every vertex in $R_{\delta_0(v)-k} \cap A$ is in distance
at most $2\delta$ from $\gamma_k(v)$, we have $\delta_N \leq 4\delta+1$. In particular, the check in the last paragraph needs to be only checked for $m \leq \delta_N/2$ (i.e., $m\leq 2\delta$).

Also, since every vertex in $R_{\delta_0(v)-k} \cap A$ is in distance at most $2\delta$ from $\gamma_k(v)$, the number of segments that include the given element, or are adjacent 
to the given element, is bounded in a bounded degree graph. Therefore, the number of segments $S'$ such that $(S,S') \in N$ is bounded.

To show efficiency, note that if $S=(A,h,j)=P^k(v)$ for some $v \in V$, the following hold:
\begin{itemize}
\item $A$, $h$, $j$ are of correct types,
\item $A$ contains exactly one leading vertex, i.e., $v \in A \cap S^{\delta_0(S)}$ such
that $h(v) = 0$; this leading vertex corresponds to $\gamma_k(v)$,
\item All the other vertices in $A$ are in bounded distance from the leading vertex (as mentioned above), 
\item The functions $h$ and $j$ are bounded (as mentioned above).
\end{itemize}

For an effective representation we consider all the potential segments $S = (A,h,j)$, which are not necessarily of form $P^k(v)$ for some $v \in V$, but all the conditions above hold.
Since the number of potential segments with the given leading vertex is bounded by a fixed constant, this redundancy does not harm the time complexity.
$P$ is a well-defined (partial) function for the potential segments, thus we can generate all the children of a given $S$ by generating all the potential segments $S'$ whose
leading vertices are the children of the leading vertex of $S$, and filtering out those for which $P(S') = S$ does not hold.
\end{itemize}
\end{proof}

\section{DHRG applications}\label{sec:dhrg_app}

Fix a RGHT $G$, parameters $n$, $R$, a probability distribution $X$ on $\{0,\ldots,R\}$,
and $p: \{0,\ldots,2R\} \ra [0,1]$. We generate a Discrete Hyperbolic Random Graph (DHRG) $H = (V,E)$ as follows:

\begin{itemize}
\item For every vertex $v \in V = \{1, \ldots, n\}$ we randomly choose $r$ according to $X$,
and a point $\mu(v) \in R^r(G)$;
\item Every pair of vertices $v_1, v_2 \in V$ is connected with an edge with probability
$p(\dist(\mu(v_1), \mu(v_2)))$.
\end{itemize}

\subsection{Choosing the parameters}
Let $a(r)$ be the probability of choosing a specific vertex $v$ such that $\delta_0(v)=r$; we have
$a(r) = P(X=r)/|R_r(G)|$. Then we can compute approximate degree distribution and clustering
coefficient as follows:                                          

\begin{itemize}
\item The average degree of $H$ can be computed using Theorem \ref{thm:stgd}.
Take the template graph $C = \{\{c_1,c_2\}, \{\{c_1,c_2\}\}\}$, both vertices of the default color $k_0$.
Then the average degree is: \[\sum_d \Count(d) \cdot a(d(c_1)) \cdot N \cdot a(d(c_2)) p(d(c_1,c_2)).\]

\item The approximate degree distribution of $H$ can be computed similarly:
the number of vertices at distance $r$ is $N \cdot P(X=r)$, and their average degree is
\[\sum_d [d(c_1)=r] \Count(d) \cdot N \cdot a(d(c_2)) \cdot p(d(c_1,c_2)) / |R_r(G)|.\]
Various vertices in distance $r$ will have different expected degrees depending on their exact
placement, but this computation gives an approximation.

\item The clustering coefficient of a graph $(H,E)$ is the probability that, for three 
randomly chosen vertices $v_1$, $v_2$ and $v_3$, we have $\{v_2, v_3\} \in E$, under the condition
$A$ that $\{v_1, v_2\}, \{v_2, v_3\} \in E$. Let the template graph $C$ be a triangle with vertices $c_1$, $c_2$, $c_3$.
The probability of our condition is
\[P(A) = \sum_d \Count(d) \cdot a(d(c_1)) a(d(c_2)) a(d(c_3))\cdot p(d(c_1,c_2)) p(d(c_2,c_3))\]
while the probability of $A$ and $\{v_2, v_3\} \in E$ ($A\cap B$) is
\[P(B\cap A) = \sum_d \Count(d) \cdot a(d(c_1)) a(d(c_2)) a(d(c_3)) \cdot p(d(c_1,c_2)) p(d(c_2,c_3)) p(d(c_1,c_3))\]
By dividing these two values, we get an approximation of the average clustering coefficient
of a DHRG.
\end{itemize}

\subsection{Computing the log-likelihood and local search}
Let $m:V \ra G$ be a DRHG embedding of $H = (V,E)$, where $|V|=n$, $|E|=m$.
Let $C = \{\{c_1,c_2\}, \{\{c_1,c_2\}\}\}$, both vertices of the single color $k$.
Color every $\mu(v)$ with color $k$; we can use Theorem \ref{thm:stg} to obtain the
number of pair of vertices in distance $d$, for every $d$, in time $O(n R^3)$;
we can also do this in $O(nR^2)$ (see Theorem \ref{thm:stgg} below). 
We can also compute $\mu(h(v_1), h(v_2))$ for every $\{v_1, v_2\} \in E$, and thus
for every distance $d$, we know how many pairs of vertices in distance $d$ there
are, and how many of them are edges. Computing the log-likelihood is straightforward.
We can also easily recompute the log-likelihood after moving a vertex of degree $a$
in $O(R^2+aR)$.

\subsection{Generating a DHRG}
To generate a DHRG in time $O(nR^2+mR)$, first place all the vertices according to the model.
Then, for every vertex $v$ and for every $d$, compute the number $q$ of other vertices $w$ such that
$\dist(v,w) = d$, using the algorithm from Theorem \ref{thm:stgg}. Each of these $q$ vertices $(w_1, w_2, \ldots, w_q)$ will create an
edge $(v,w)$ with probability $p(d)$. We choose $w_{X_1}$, $w_{X_1+X_2}$, \ldots, where
$X_i$ has the geometric distribution with parameter $p(d)$, until the index exceeds $q$. We need
to modify the algorithm from Theorem \ref{thm:stgg} to find out the actual vertex with the given index,
which is straightforward.

\begin{theorem}\label{thm:stgg}
Fix an efficient segment tree graph $\calS$. Let $R \in \bbN$. 
Let $B_R = \bigcup_{d \leq R} P^{-d}(v_0) \cap V_\calS$ be the ball of radius $R$ in $V_\calS$.
Then there exists a data structure with the following operations:

\begin{itemize} 
\item $\InitCounter$, which initializes $\val: B_R \ra \bbR$ to 0 for every $k \in K$,
\item $\Add(v, x)$, which adds $x$ to $\val(v)$, where $v \in B_R$,
\item $\Count(d)$, which for $d \in [0..2R]$ returns $\sum_{v_1} \sum_{v_2} \val(v_1) \val(v_2) [\delta(v_1,v_2)=d]$,
where $[\phi]$ is 1 iff $\phi$ is true and 0 otherwise.
\end{itemize}

Such $\Count$ and $\InitCounter$ can be implemented in $O(1)$, and $\Add$ can be implemented in $O(R^2)$.
\end{theorem}

\begin{proof}
For every $s \in \calS$, let $c(s,d) = \sum{v \in P^{-*}(s)} \val(v) [\delta_0(v)=d]$. 
We maintain the current values of $c$ for every $s \in \calS$ and $d \in [0,\ldots,R]$.
We also maintain an array of precomputed results of $\Count$.

After each $\Add(v,x)$ operation, we update $c$ for every ancestor $s$ of $x$, and we also
update $\Count$. Updating $c$ is straightforward; $\Count(d)$ will change by
\[\sum_{d_1=0..\delta_0(v)} \sum_{s' \in N(P^d(v))}  c(s',d-(d_1-\delta_0(v))-\delta_N(s,s')).\]
\end{proof}

\nonblind{\section{Visualization}
See \url{http://www.mimuw.edu.pl/~erykk/segviz/index.html} for a visualization of some grids in the hyperbolic plane.}

\end{document}